\documentclass[iop,appendixfloats]{emulateapj}

\newcommand{\figurepath}{.}

\usepackage{amsmath}
\usepackage{natbib}
\usepackage{color}
\usepackage{multirow}
\usepackage{graphicx}
\definecolor{orange}{rgb}{1,0.5,0}
\usepackage[colorlinks=True,citecolor=orange,urlcolor=blue,linkcolor=red]{hyperref}

\long\def\symbolfootnote[#1]#2{\begingroup%
\def\thefootnote{\fnsymbol{footnote}}\footnote[#1]{#2}\endgroup}

\newcommand{\msun}{{\rm M}_{\sun}}

\newcommand{\rsun}{{\rm R}_{\sun}}

\newcommand{\rstar}{{\rm R}_{\star}}

\begin{document}
\title{Jet formation from massive young stars: \\
       Magnetohydrodynamics versus radiation pressure}
\shorttitle{MHD jets and radiation pressure}
\author{Bhargav Vaidya\altaffilmark{1}}
\author{Christian Fendt}
\author{Henrik Beuther}
\author{Oliver Porth\altaffilmark{1}}
\affil{Max Planck Institute for Astronomy, K\"onigstuhl 17, D-69117 Heidelberg, Germany} 
\email{Email: vaidya@mpia.de, fendt@mpia.de}
\altaffiltext{1}{Member of the \textit{International Max Planck Research School for 
              Astronomy and Cosmic Physics at the University of Heidelberg} (IMPRS-HD), and the
              \textit{Heidelberg Graduate School of Fundamental Physics} (HGSFP)}
\begin{abstract}
Observations indicate that outflows from massive young stars are more collimated during their early 
evolution compared to later stages.
Our paper investigates various physical processes that impacts the 
outflow dynamics, i.e. its acceleration and collimation.
We perform axisymmetric MHD simulations particularly considering the radiation 
pressure exerted by the star and the disk.
We have modified the PLUTO code to include radiative forces 
in the line-driving approximation.
We launch the outflow from the innermost disk region ($r < 50$\,AU)
by magneto-centrifugal acceleration.
In order to disentangle MHD effects from radiative forces, we start the 
simulation in pure MHD, and later switch on the radiation force.
We perform a parameter study considering different stellar masses (thus luminosity),
magnetic flux, and line-force strength.
For our reference simulation - assuming a $30\msun$ star, we find substantial de-collimation of 35\%
due to radiation forces. 
The opening angle increases from $20^\circ$ to $32^\circ$ for stellar masses from
$20\msun$ to $60\msun$.
A small change in the line-force parameter $\alpha$ from 0.60 to 0.55 changes the opening angle 
by $\sim 8^\circ$. 
We find that it is mainly the stellar radiation which affects the jet 
dynamics. Unless the disk extends very close to the star, its is too small to have much impact.
Essentially, our parameter runs with different stellar mass can be understood 
as a proxy for the time evolution of the star-outflow system.
Thus, we have shown that when the stellar mass (thus luminosity) increases
(with age), the outflows become less collimated.   
\end{abstract}


%
%
\section{Introduction}
\label{sec:Intro}
Outflows and jets are integral processes of star formation.
They are believed to be essential for the angular momentum evolution of 
the cloud core and the protostar - either directly as stellar winds or indirectly 
by changing the structure and the evolution of the surrounding accretion disk.
Also, outflows from young stars provide an important feedback mechanism to return
mass and energy into the ambient medium from which the young star is born.  

Most of the current understanding about the formation and propagation of jets/outflows 
comes from observations of low-mass stars. 
In this case we are fortunate to know the leading dynamical parameters such as outflow 
velocity, density, and temperature (e.g. \citealt{Hartigan:2007p2295}, \citealt{Ray:2007p3306}). 
However, in case of massive young stars many of these 
  parameters are poorly known, 
although large multi-wavelength studies for massive star forming regions have been done
over the last decade 
(\citealt{Beuther:2002p3574}, \citealt{Stanke:2002p4352}, \citealt{Zhang:2005p4269}, \citealt{LopezSepulcre:2009p4143}, \citealt{LopezSepulcre:2010p4271}, 
\citealt{Torrelles:2011p4365}). 
These studies suggest that outflows are an ubiquitous phenomenon not only for low-mass 
stars, but also in massive star forming regions.

The standard framework for the launching process of jets or outflows from low-mass protostars
(and most probably also for extragalactic jets) is the model of a disk wind accelerated and
collimated by magneto-centrifugal and magnetohydrodynamic forces 
(\citealt{Blandford:1982p892}, \citealt{Pudritz:2007p727}).
A number of numerical simulations of jet formation have been performed which all confirm this picture of self-collimated MHD jets for low mass stars
(\citealt{Ouyed:1997p634}, \citealt{Krasnopolsky:1999p577}, \citealt{Fendt:2002p1135}, \citealt{Ouyed:2003p1120}, \citealt{Fendt:2006p574}, \citealt{Fendt:2009p710}).
Its applicability has also been demonstrated for the case of extragalactic jets (e.g.\citealt{Komissarov:2007p3383},  \citealt{Porth:2010p1041}).

The formation of the jets around young, still forming high-mass stars takes place in the deeply
embedded cold dust and gas cocoons exhibiting large visual extinction of the order 
100 to 1000\,mag \citep{Arce:2007p798}. 
Recent progress in observations of outflows from young high-mass stars comes from mm and cm
wavelengths. 
Early low-spatial-resolution single-dish studies suggested that massive outflows may have a 
lower collimation degree than those of their low-mass counterparts
\citep{Shepherd:1996p3451}. 
Also, one of the famous outflow, the Orion-KL system, exhibits a more chaotic and not 
collimated structure \citep[e.g.,][]{Schulz:1995p3570}. 
However, later studies indicated that the collimation degree of jets from high-mass
protostars can be as high as from low-mass regions (e.g. \citealt{Beuther:2002p3574}, \citealt{Beuther:2002p1149}, \citealt{Gibb:2003p3603}, \citealt{Garay:2003p3647}, \citealt{Brooks:2003p3687}, \citealt{Beuther:2004p3588}, \citealt{Davis:2004p3719}). 
Typical outflow rates in high-mass systems are estimated to be a few times $10^{-5}$
to $10^{-3}$\,M$_{\odot}$\,yr$^{-1}$, 
implying accretion rates on the same order of magnitude (e.g.,\citealt{Beuther:2002p1149}).

The originally puzzling result that different studies found different degrees of jet collimation
for high-mass star-forming regions could qualitatively be resolved when it was realized that these 
studies in fact targeted different evolutionary stages. 
While jets found in the youngest star forming regions appear similarly collimated as their 
low-mass counterparts,
jets from more evolved (massive) stars exhibit much broader outflow cones. 
To account for this effect \cite{Beuther:2005p791} proposed an evolutionary picture for high-mass 
outflows, where the jet formation starts with similar MHD acceleration processes compared to the 
low-mass models during the earliest evolutionary stages. 
However, as soon as the central sources gain significant mass, other processes come into play, 
for example, the radiation from the central star or more turbulence at the base of the jet. 
Any combination of this and other processes was expected to lower the degree of collimation. 
However, the evolutionary sequence proposed by \cite{Beuther:2005p791} was largely observationally
motivated and could only provide a qualitative explanation.
A thorough theoretical treatment to account for this observational result is still 
missing - a topic which we now address in the present paper.

How the radiation field affects the formation of a jet is not obvious {\it a priori}.
In order to quantify and to disentangle the physical processes involved, a detailed numerical 
investigation is required. 
Essentially, stellar (and disk) radiative forces may affect jet acceleration and collimation 
directly (neglecting ionization, heating, and probably turbulent stirring for simplicity), 
but also in-directly by changing the physical conditions of the jet launching area, 
thus governing the mass loading or the initial entropy of the ejected jet material.
For example, numerical MHD simulations have shown that jets with higher (turbulent) magnetic 
diffusivity are expected to be substantially less collimated (\citealt{Fendt:2002p1135}).

Our previous studies \citep{Vaidya:2009p1104} have shown that the inner accretion disk around 
massive protostars is ionized, has a high temperature and  is  gravitationally and thermally 
sufficiently stable in order to provide a suitable launching  area  for an outflow.
Together with the  recent observations of  magnetic fields around high
mass protostars (e.g\,\citealt{Vlemmings:2008p596, Vlemmings:2010p2895},\citealt{Girart:2009p808}) , this  indeed 
supports the picture of a scaled-up version of low-mass stellar jet formation.

In this paper we will present a detailed  investigation of how a strong radiation field impacts 
the structure and dynamics of a magneto-hydrodynamical driven jet. 
Motivated  by the presence of strong jets and outflows in massive star formation, we apply the 
standard picture of MHD jet formation known for  astrophysical jets and put it in the physical 
environment of a massive  young star.

%
%

\section{Model setup: Jet formation from massive young stars}
\label{sec:modelsetup}
\begin{figure} \centering
\includegraphics[width=9.0cm]{\figurepath/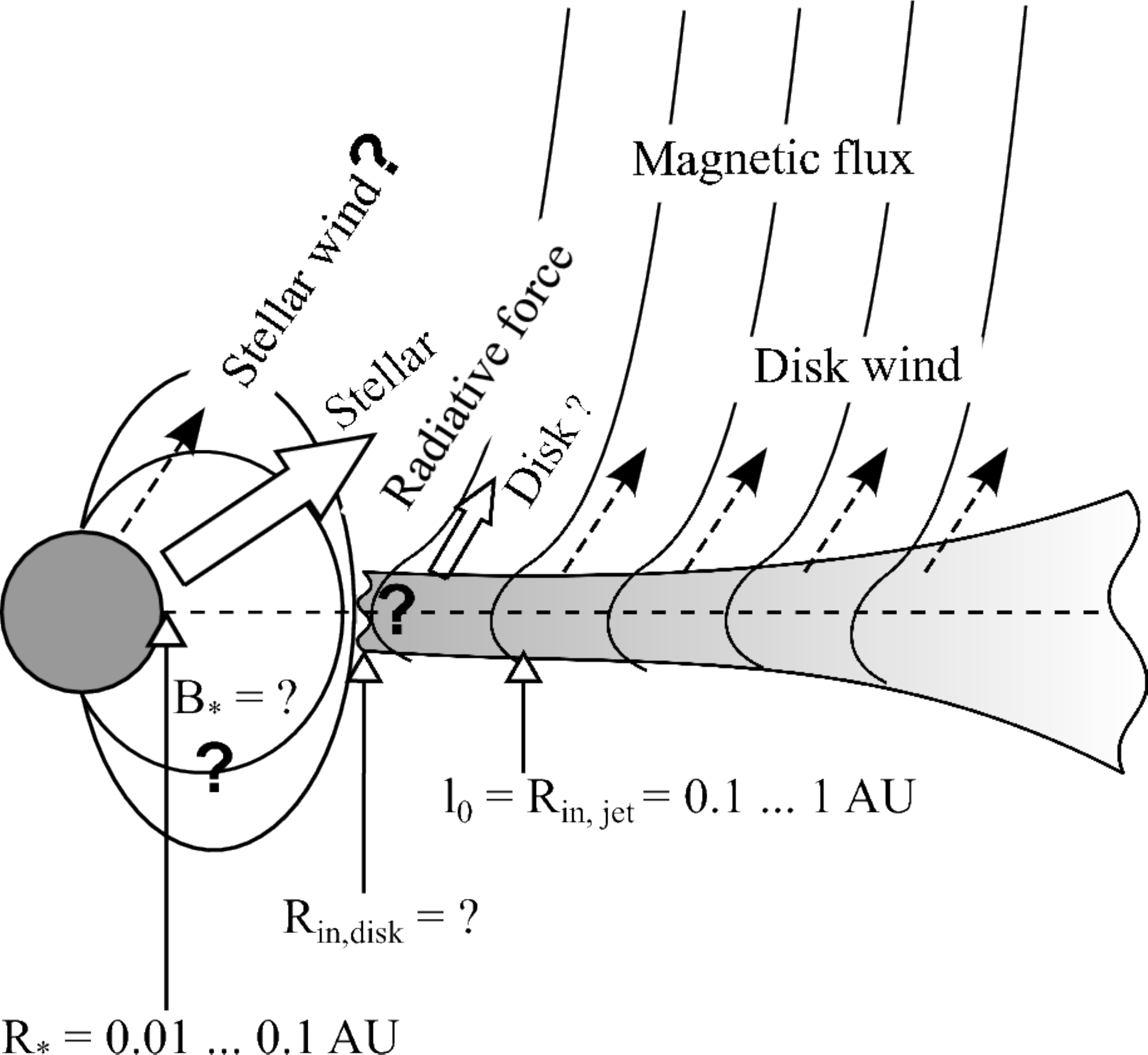}
\caption{Sketch showing our model setup of inner regions around a
 massive young star. 
 Several constituents are considered: 
 The disk outflow (\textit{dashed arrows}) is launched along the magnetic flux
 surfaces. 
 The inner most launching point is denoted by $R_{\rm in,jet}$. 
 This outflow is subject to radiation forces (\textit{white arrow}) from the 
 star and (potentially) from an inner hot accretion disk. 
 The stellar radius $R_{*}$, could be as large as $\sim 100\,\rsun$, 
 while the stellar magnetic field structure and strength $B_{*}$ is 
 rather uncertain. 
 The location of an inner disk radius $R_{\rm in,disk}$ (if existent) 
 is also not known.}\label{physetup}
\end{figure}

In this section we discuss the model setup applied for our numerical study 
concerning the formation of jets and outflows from massive young stars.
The central point is that we are going to consider the main features of the 
{\em standard model} of MHD jet formation which is well established for 
low-mass young stars or Active Galactic Nuclei also for high-mass young 
stars.
Our model consists of the following essential ingredients
(see Fig.~\ref{physetup}).

\begin{itemize}
\item A central massive young star which is rapidly evolving in
      mass $M_{*}$, luminosity $L_{*}$, and radius $R_{*}$.
\item A surrounding accretion disk with high accretion rate, estimated
      to be of the order of $\dot{M} \simeq 10^{-3}\,\msun\,\rm yr^{-1}$.
\item A jet launching inner accretion disk. 
      The extension of this area towards the star is not known and may 
      depend on the existence of a strong stellar magnetic field and 
      stellar radiation pressure. 
      Instead of introducing an inner disk radius $R_{\rm in, disk}$, 
      we will refer to an
      inner jet launching radius $R_{\rm in, jet}$, 
      which we presume to be between 0.1 and 1.0 AU, and to which
      the length scale of the simulation will be normalized $l_0 =R_{\rm in, jet}$. 
\item A magnetic field around the protostellar
      object. Magnetic fields are essential for generating collimated 
      high-speed outflows. Observational indication for such fields 
      around high-mass protostars exists.
\item A strong radiation field of the high-mass young star which may influence
      accretion and ejection processes. 
      In case when the accretion disk reaches down to radii close to the 
      stellar surface (no gap as in case of low mass stars), also the high 
      disk luminosity may play a role for the outflow dynamics.
\end{itemize}

In the following we briefly discuss the observational and theoretical
background of these constituents
and finally mention the limits of our model and possible
model extensions (a more detailed discussion is provided before the summary).

\subsection{The central massive young star}
\label{ssec:star}
It has been proposed that outflows from massive young stars may follow an evolutionary sequence
such that outflows tend to be more collimated and similar to jets from low-mass stars in the 
early stages of stellar evolution, whereas at later times the outflows are less collimated \citep{Beuther:2005p791}.
Since this intrinsically corresponds to an evolutionary sequence in stellar mass,
we have investigated simulations with different central mass, ranging from 
$20\,\msun$  to $60\,\msun$. 

A higher stellar mass automatically gives rise to a faster outflow (supposed the
relative launching radius is the same), just because the outflow originates deeper
in the gravitational well.

\subsection{The accretion disk and the jet launching area}
\label{ssec:jetregion}
Jets from low mass stars are thought to be launched from less than 1\,AU of
the inner accretion disk (e.g.\,\citealt{Anderson:2003p1140}, \citealt{Ray:2007p3306}).
In T Tauri stars, these regions could be well studied via NIR interferometry
(e.g.\, \citealt{Akeson:2000p6781, Akeson:2005p6709},
\citealt{Dullemond:2010p5781}). 
However, to probe regions $\leq$ 100 AU around a young high-mass star is difficult 
due to the large extinction. 
We had therefore studied this region via semi-analytic modeling in a previous 
paper \citep{Vaidya:2009p1104} to get handle on the physical properties of such 
disks.
We have shown that here the disk may reach high temperatures $\sim 10^{5} \rm{K}$ 
leading to sublimation of most of the dust and ionizing the bulk of the material.

The inner disk radius in case of low mass stars is usually estimated assuming
magnetic pressure balance with the accretion ram pressure. 
Typical values are $\sim 3-5\, \rstar$. 
For young massive stars, such an estimate is not possible due to lack 
of knowledge on stellar magnetic fields during the formation stage. 
In addition, radiative force from the bright luminous massive star could 
also influence dynamics of the inner disk. 
Although our semi-analytic modeling indicated that the disk could extend 
right down to the central star, detailed 3D models with accurate radiative 
treatment are required to get more insight in these close-by regions 
(Vaidya et al. {\em{in prep}}). 

We have chosen the inner launching point at a distance of $l_0 = 1$\,AU 
from the star (i.e. similar to low mass stars).
A value of $l_0 < 0.1$\, AU would imply a high rotation speed and
a deep potential well, resulting in a faster outflow.
At the same time the jet launching part of the disk would be much hotter
and possibly result in a higher contribution to the radiation force
from these hot inner parts. 
On the other extreme, large jet launching radii $l_0 > 10$\,AU will
result in slow outflows, barely affected by stellar and disk radiation forces.

The density $\rho_0$ at the inner launching point is used for a
physical scaling. 
We estimate $\rho_0$ from the observed mass fluxes, which are typically of 
the order of $10^{-3} - 10^{-5}\msun$\, yr$^{-1}$, providing
$\rho_0 \simeq 10^{-13} - 10^{-15}$ $\rm{g\, cm}^{-3}$ (\S~\ref{sec:rhobase}). 

\subsection{The magnetic field}
\label{ssec:jetregion}

For many years the role of magnetic fields in massive star formation was 
not really known. 
However, recent observations have detected relatively 
strong magnetic fields in massive star forming regions \citep{Vlemmings:2008p596}. 
Polarimetric  observations of the hot massive molecular core HMC G31.41 have revealed 
a large-scale hourglass-shaped magnetic field configuration \citep{Girart:2009p808}.
\citet{Beuther:2010p4089} detected a magnetic field aligned with
the molecular outflow via polarimetric CO emission. 
Further observations have detected synchrotron emission from the 
proto-stellar jet HH\,80-81, indicating a $\sim 0.2$\,mG magnetic field in the 
jet knots while the stellar mass of $\sim 10\,\msun$ is in the range of 
massive stars \citep{CarrascoGonzalez:2010p3410}.

\cite{Vlemmings:2010p2895} have measured magnetic field strengths using polarization 
by 6.7\,GHz methanol masers around the massive protostar Cepheus A HW2 and 
derive a line of sight (l.o.s) magnetic field strength $\sim 23$\,mG at a distance of 300-500\,AU from 
the central star. 
The magnetic field strength is parametrized
by the plasma beta, $\beta_0$, which is the ratio of the thermal gas
pressure to the magnetic pressure at the inner launching point $l_0$.
Our simulations are so far limited to $1< \beta_0 <10$ by numerical and physical
reasons. 
This translates into field strengths at 1\,AU of $\sim$ 100 times weaker than estimated
by conserving magnetic flux using observed values at 300\,AU. 

Note that it is not only the field strength but also the field {\em distribution} which 
affects the jet formation process, as it was shown by (\citealt{Fendt:2006p574})
by MHD jet formation considering a wide parameter set of magnetic field and mass 
flux distribution along the jet launching area.
As a result, simulations applying a concentrated magnetic flux profile tend to be 
less collimated. 
We apply as initial field distribution the standard potential field 
suggested by \cite{Ouyed:1997p634} (\S~\ref{ssec:Initcond}). 
A central stellar dipolar field (see Fig.~\ref{physetup}) is not (yet) supported
by observations.

\subsection{The stellar and  disk luminosity}
The massive young star produces a substantial luminosity, which is supposed to
dynamically change the outflow structure.
The dependence of the radiation force on the stellar luminosity is parametrized by the Eddington ratio
$\Gamma_{\rm e}$, defined as the ratio of the radiation force due to
electron scattering to the central gravity (see Table ~\ref{ParaSummary}). 
The characteristic values of this dimensionless parameter are obtained from the stellar
evolution model of \citet{Hosokawa:2009p4005}.
In order to study the impact of radiation forces on the dynamics of outflows, we first 
launch a collimated jet from a disk wind via MHD forces. 
Then, after the pure MHD jet has achieved a steady state, we initiate the radiation 
forces, and then compare their impact on the MHD jet.
Also, simulations in which the radiation forces are considered from the beginning
(which are computationally much more expensive) end up with a final
structure of the outflow  similar to the one obtained from the step-by-step method. 

We prescribe the radiation force by the  line-driving mechanism introduced by
Castor, Abbott, \& Klein (CAK). 
Such a force is parameterized by two physical parameters $k$ and
$\alpha$. 
The value of $k$ is proportional to the total number of lines. 
The quantity $\alpha$ can be considered as a measure for the ratio of acceleration from 
optically thick lines to the total acceleration \citep{Puls:2000p2903}.
Depending on the selection of lines, for  massive OB stars typical values obtained for $k$ range
from 0.4-0.6, while $\alpha$ ranges  between 0.3 and 0.7 \citep{Abbott:1982p900}.
The process of line driving has also been applied to cataclysmic variables (CVs) 
(\citealt{Feldmeier:1999p856}), as well as to hot and luminous disks around Active 
Galactic Nuclei (AGN) (\citealt{Proga:2000p853}, \citealt{Proga:2004p838}). 

\cite{Proga:2003p1132} carried out numerical simulations driving winds from hot luminous magnetized accretion disks of AGNs, assuming spherical geometry, an isothermal equation of state, and
an initially vertical magnetic field structure.
Here we consider a potential field which is hour-glass shaped, and 
an adiabatic equation of state.

\subsection{Limitations of our model setup}
Our paper, for the first time, provides a quantitative study of the interplay 
between radiative and MHD forces on outflows launched from the vicinity of 
young massive stars, applying high-resolution axisymmetric numerical simulations.
However, a few critical points can be raised which may limit the applicability
of our model and which should probably be considered in forthcoming investigations.

From the general point of view there is the lack of true knowledge concerning a 
number of important parameters as discussed above.
One important question is the location of the inner jet launching radius.
If it is identical with an inner disk radius - where is that inner disk radius
located, if it exists at all? Is there a strong stellar magnetic field which could 
open up a gap between the stellar surface and the disk as it is 
known for low-mass young stars?

Further, our disk model is taken as a boundary condition steady in time.
As the star evolves, also the disk structure and accretion rate may evolve
in time. This question could only be answered by simulations
solving also for the disk structure. We do not explicitly incorporate
heating and cooling but simply consider the adiabatic expansion and
compression in the jet.

Another question is the existence of a stellar wind.
We know that OB stars have strong mass loss in form of stellar winds during
{\em{later stages}} of their life times. 
The derived mass loss rates are typically of the order of
$10^{-6}\,\msun\,{\rm yr}^{-1}$. 
These winds are primarily radiation driven via the line-driving mechanism
(e.g. \citealt{Kudritzki:2000p2061}, \citealt{Owocki:2009p2068}).
The velocities derived are high, ranging from $\sim 500-1000\,\rm{km}\,\rm{s}^{-1}$ 
and are usually supersonic.
However, in case of high-mass {\em young} stars, no indication for such winds
has been found so far possibly due to high obscuration.
Nevertheless, a future study should implement the physical effect of a 
central stellar wind.

%
%
\section{Basic equations}
\label{sec:eqns}

For our study, we carry out axisymmetric numerical ideal MHD simulations using the PLUTO 
code \citep{Mignone:2007p644}.
We have modified the original code to incorporate source terms treating the line-driven forces
from central star and disk, taking into account self-consistently the density and velocity
distribution of the outflow.
  
The MHD code considers the following set of equations.
That is the conservation of the mass, momentum, and energy,
\begin{equation}\label{masscons}
\frac{\partial \rho}{\partial t} + (\vec{v} \cdot \nabla)\rho  +
\rho \nabla \cdot \vec{v} = 0,
\end{equation}
\begin{eqnarray}\label{momcons}
\rho(\frac{\partial \vec{v}}{\partial t} +
(\vec{v} \cdot \nabla) \vec{v}) = ~~~~~~~~~~~~~~~~~~~~~~~ \nonumber \\
- \nabla P + \frac{1}{4\pi} (\nabla \times \vec{B}) \times \vec{B}
- \rho \nabla \Phi + \rho \vec{F}^{\rm rad},
\end{eqnarray}
\begin{eqnarray}\label{encons1}
\frac{\partial}{\partial t}(\rho E)
+ \nabla \cdot\left[ \rho E \vec{v} + (P + \frac{B^2}{8\pi})\vec{v}\right]  
- \vec{B}(\vec{v}\cdot\vec{B}) = ~~ \nonumber \\
 \rho\left[-\nabla\Phi + \vec{F}^{\rm rad}\right] \cdot \vec{v} \quad,
\end{eqnarray}
where $\rho$ is the gas density, $\vec{v}$ the velocity vector, 
$P$ the gas pressure, and $\vec{B}$ the magnetic field vector with the 
poloidal and toroidal components $\vec{B}_{\rm p}, {B}_{\phi}$.
In order to include the radiative  forces $\vec{F}^{\rm rad}$, 
the relevant  source terms have been added in the momentum
and energy conservation equation.
The total energy density of the flow $E$ comprises contributions from 
the internal  energy $\epsilon$, the mechanical energy, and the magnetic 
energy,
\begin{equation}\label{encons2}
 E = \epsilon + \frac{v^2}{2} + \frac{B^2}{8 \pi \rho}.
\end{equation}

The gas pressure in the flow is related to the density assuming an 
adiabatic equation 
of state with the adiabatic index $\gamma$,
\begin{equation}\label{EOS}
P = (\gamma - 1) \rho \epsilon.
\end{equation}

The evolution of the magnetic field is governed by the induction equation,
\begin{equation}\label{induction}
\frac{\partial \vec{B}}{\partial t} = \nabla \times \left(\vec{v}\times \vec{B}\right).
\end{equation}
We treat the ideal MHD equations without considering resistive terms.

In addition to the above set of equations the code obeys the condition of 
divergence-free magnetic fields, $\nabla \cdot \vec{B} = 0$, 
using the constraint transport method.

\subsection{Prescription of radiation forces}
\label{ssec:radforces}

We do not explicitly consider radiative transfer, however, we study the 
effects of momentum transfer by radiative forces on the outflow matter which is launched by 
MHD processes from the underlying disk. 
The  total radiative force $\vec{F}^{\rm rad}$ comprises of four contributions - the 
acceleration due to continuum radiation from 
star $\vec{f}_{\rm cont,*}$ and disk $\vec{f}_{\rm cont,disk}$, respectively, and, 
similarly, due to spectral lines from 
star $\vec{f}_{\rm  line,*}$ and disk $\vec{f}_{\rm line,disk}$, respectively, 
\begin{equation}\label{radforce}
\vec{F}^{\rm rad} = \vec{f}_{\rm cont,*} + \vec{f}_{\rm cont,disk} 
                  + \vec{f}_{\rm line,*} + \vec{f}_{\rm line,disk}.
\end{equation}
In case of young massive stars, the stellar radiation is sub-Eddington. 
This immediately implies that the continuum force do not contribute in modifying the 
dynamics of the MHD outflow as its strength is much below the gravitational pull of 
the central star. 
However, line forces could prove to be efficient in substantially enhancing the continuum 
force by a so-called force multiplier, which is subject to complex theoretical studies
of radiative transfer. This theory was developed first by \cite{Castor:1975p898} who 
solved the radiative transfer 
equations from spectral lines in moving atmospheres. 
According to their studies, the force due to line driving can 
be expressed as a product of force due to continuum radiation and a force multiplier.

The force multiplier depends on two parameters - $k$ and
$\alpha$. A "general" parametrization for the force multiplier which is
independent of ion thermal velocities  
$v_{\rm th}$ (see Eq~\ref{eq:tform} in the Appendix) has been introduced by \cite{Gayley:1995p1968}. 
In his formulation, the parameter $k$, initially introduced by \cite{Castor:1975p898}, has been 
replaced by a parameter $\bar{Q}$
and the force multiplier can be expressed as follows, 
\begin{equation}\label{lfGayley}
M(\mathcal{T}) = k\mathcal{T}^{-\alpha}=
\left[\frac{\bar{Q}^{1-\alpha}}{1-\alpha}\left(\frac{{|\hat{n}\cdot\nabla (\hat{n}\cdot\vec{v})|}}{\sigma_{\rm e}c\rho}\right)^{\alpha}\right],
\end{equation}
where $\mathcal{T}$ is the optical depth parameter that depends on the
l.o.s velocity gradients (see eq.~\eqref{eq:tform}).
The "strongest form" of this parameterization requires the ansatz $\bar{Q}=Q_0$, 
where $Q_0$ being  the line strength of the strongest line.

Typically for evolved
massive stars, the parameter $\bar{Q}$ lies in the range of 1000-2000 \citep{Gayley:1995p1968}. 
For our present study, we have applied the force multiplier considering this "general" 
parameterization in its "strongest form". 
Thus, we have the two parameters $Q_0$ and $\alpha$ which define the
force multiplier (Eq.~\ref{lfGayley}) and 
hence the line forces. 
For simplicity, we assume constant values for these line force parameters for 
a particular simulation run.
One of the important properties of the force multiplier $M(\mathcal{T})$ is that its values saturates 
to a maximum value $M_{\rm{max}}$ when the medium becomes optically thin and the optical 
depth parameter approaches a minimum value, $\mathcal{T} \leq \mathcal{T}_{\rm{min}}$. 
The typical value for the maximum force multiplier is relatively constant,
$M_{\rm{max}} \sim 10^{3}$, depending only on the metallicity \citep{Gayley:1995p1968}.

The line force due to the central star is a product of the force due to continuum from the 
star and  the force multiplier,
\begin{equation}\label{flinestar}
\vec{f}_{\rm line,*} = \vec{f}_{\rm cont,*}M(\mathcal{T}).
\end{equation}
The most difficult part for the numerical realization of the line forces is to calculate the proper 
l.o.s velocity gradients ${|\hat{n}\cdot\nabla (\hat{n}\cdot\vec{v})|}$ that appears in the 
formulation of force multiplier (Eq~\ref{lfGayley}). 
Following the definition of this term by \cite{Rybicki:1978p897}, we express this 
gradient in terms of the rate-of-strain tensor $e_{ij}$,
\begin{equation}\label{rateofstrain}
\hat{n}\cdot\nabla (\hat{n}\cdot\vec{v}) = \displaystyle\sum\limits_{i,j} e_{\rm ij}n_{\rm i}n_{\rm j} = \displaystyle\sum\limits_{i,j} \frac{1}{2}\left(\frac{\partial v_{\rm i}}{\partial r_{\rm j}} + \frac{\partial v_{\rm j}}{\partial r_{\rm i}}\right)n_{\rm i}n_{\rm j},
\end{equation}
where in general $v_{\rm i}$,$r_{\rm i}$, and $n_{\rm i}$ are the components of velocity
$\vec{v}$, distance $\vec{r}$, and the unit vector $\hat{n}$, respectively.
The different components of this tensor in cylindrical
coordinates were calculated from \cite{Batchelor:mine}.
\cite{Proga:1998p854} approximated the above sum to be equal to the most dominant 
term, corresponding to the radial gradient of flow velocity along the
spherical radius. Simulations with a more accurate algorithm of including all the other terms to determine the l.o.s. velocity
gradient using the above equation have also been carried out
\citep{Proga:1999p872}. These simulations show that the qualitative
features of winds are not changed as compared to the approximate
calculation of the gradients. Further, they are numerically
very expensive.
We can approximate that the 
region of consideration is far away from the star as the central object is a point source. 
Therefore, the gradient can simply set to be equivalent to $dV_R/dR$, 
where $R$ is the spherical radius and $V_R$ is the gas velocity along the 
radius $R$. 
However, when using cylindrical coordinates instead of spherical coordinates we have 
to re-write equation~\eqref{rateofstrain} by transforming $R$ to $r$, and 
thereby adding further terms,
\begin{equation}\label{mygrad}
\hat{n}\cdot\nabla (\hat{n}\cdot\vec{v}) = 
\frac{1}{1 + x^2}
\left(\frac{\partial v_{r}}{\partial{r}} 
   + x\left(\frac{\partial v_{\rm r}}{\partial z} 
   + \frac{\partial v_{\rm z}}{\partial r}\right) 
   + x^2 \frac{\partial v_{\rm z}}{\partial z}\right),
\end{equation}
where $x = z/r$.

We have applied two of the line force components individually 
as source terms in order to disentangle their effects on the dynamics of a 
(pure) MHD jet launched from the underlying disk. These two components
are from the central star alone and other due to underlying hot
accretion disk. Disks around massive young stars accrete very rapidly with rates of 
$10^{-5}\msun\,\rm yr^{-1}$ to $10^{-3} \msun\,\rm yr^{-1}$. 
Such high accretion rates imply very high accretion luminosities, in particular in
their very inner regions. 
In order to calculate the line forces due to the disk luminosity, proper geometric factors 
have to be taken into account. We apply a formulation similar to
\cite{Pereyra:2000p836} whose details are given in the appendix.

%
%
\section{Numerical setup}
\label{sec:numsetup}
\subsection{Grid setup \& physical scaling}
\label{ssec:gridandscaling}
We perform axisymmetric ideal MHD simulations of jet formation 
in the presence of radiative forces. 
Our simulations are carried out on a grid of physical size 
$(r \times z) = (52\,l_0 \times 152\,l_0)$, where $l_0$ denotes the physical length scale.
The grid is divided into $(512 \times 1024)$ cells in radial and vertical direction,
respectively. 
Within $r < l_0$ and $z < l_0$ the grid is uniform with a resolution of $\delta r = \delta z = 0.05$,
while for $l_0 < r < 50\,l_0$ or $l_0 < z < 150\,l_0$ the grid is stretched with a ratio of 
1.002739 and 1.001915 in radial or vertical direction, respectively.
The remaining, very outer part of the grid is again uniform with cell size $\delta r = 0.125$ and 
$\delta z = 0.20$, respectively. 
 
Pure MHD simulations would be scale-free.
However, when radiation is considered a physical scaling of the dynamical 
variables becomes essential. Three scaling parameters in physical units are used -- the length scale
$l_0$, the base flow density $\rho_0$ and the Keplerian velocity
$v_0$,  at
the inner launching point (see \S~\ref{sec:modelsetup} for physical
values used for scaling parameters.)

All other quantities can be derived using the following definitions,
\begin{eqnarray}\label{scaledef}
 r_{\rm c} = \frac{r_{\rm cgs}}{l_0} , z_{\rm c} = \frac{z_{\rm cgs}}{l_0},  \rho_{\rm c} = \frac{\rho_{\rm cgs}}{\rho_0} , v_{\rm c} = \frac{v_{\rm cgs}}{v_0} ,  \nonumber \\
p_{\rm c} = \frac{p_{\rm cgs}}{\rho_0
v_0^{2}} ,   B_{\rm c} = \frac{B_{\rm cgs}}{B_0} = \frac{B_{\rm cgs}}{\sqrt{4\pi\rho_0 
v_0^2}}.
\end{eqnarray}

The force multiplier defined in equation~\eqref{lfGayley} can be
expressed in code units as follows
\begin{equation}\label{nondimforcemult}
M_{\rm c}(\mathcal{T_{\rm c}}) =
\left[\frac{Q_0^{1-\alpha}}{1-\alpha}\left(\frac{v_0}{\sigma_{\rm
        e}c\rho_0l_0\rho_{\rm c}}\left|\frac{dv_{\rm l}}{dl}\right|\right)^{\alpha}\right].
\end{equation}

Quantities with subscript \textit{'c'} are obtained from
simulations in the dimensionless form where as the quantities with subscript
\textit{'cgs'} are the ones required in the physical units. 
We measure the time $t_{\rm c}$ in units of ${l_0}/{v_0}$.
We denote the number of rotations at the inner launching point $l_0$ as
$N_{\rm rot} =l_0/2\pi v_0 = t_{\rm c}/2\pi$

Using these normalizations the conservation of momentum (Eq.~\ref{momcons}) can be written 
in dimensionless code units,
\begin{eqnarray}
\label{nondimmomcons}
\rho_{\rm c}\left(\frac{\partial \vec{v}_{\rm c}}{\partial t_{\rm c}} + (\vec{v}_{\rm c}\cdot \nabla_{\rm c})\vec{v}_{\rm c}\right) = \frac{2}{\beta_0}((\nabla_{\rm c} \times \vec{B}_{\rm c}) \times \vec{B}_{\rm c})   \nonumber \\
- \nabla_{\rm c} \vec{P}_{\rm c}- \rho_{\rm c} \nabla_{\rm c} \Phi_{\rm c}  + \rho_{\rm c} (\vec{F}_{c}^{\rm rad}),
\end{eqnarray}
where the plasma beta, 
$\beta_0 = (8\pi\rho_0v_0^{2})/B_0^{2}$, 
specifies the initial strength of the (poloidal) magnetic field at
$l_0$ and the radiation force 

\begin{equation}
 \vec{F}^{\rm rad}_{\rm c} = \frac{\vec{F}^{\rm rad}_{\rm cgs}}{v_0^{2}/l_0}. 
\end{equation}

\begin{figure} \centering
\includegraphics[width=10.0cm]{\figurepath/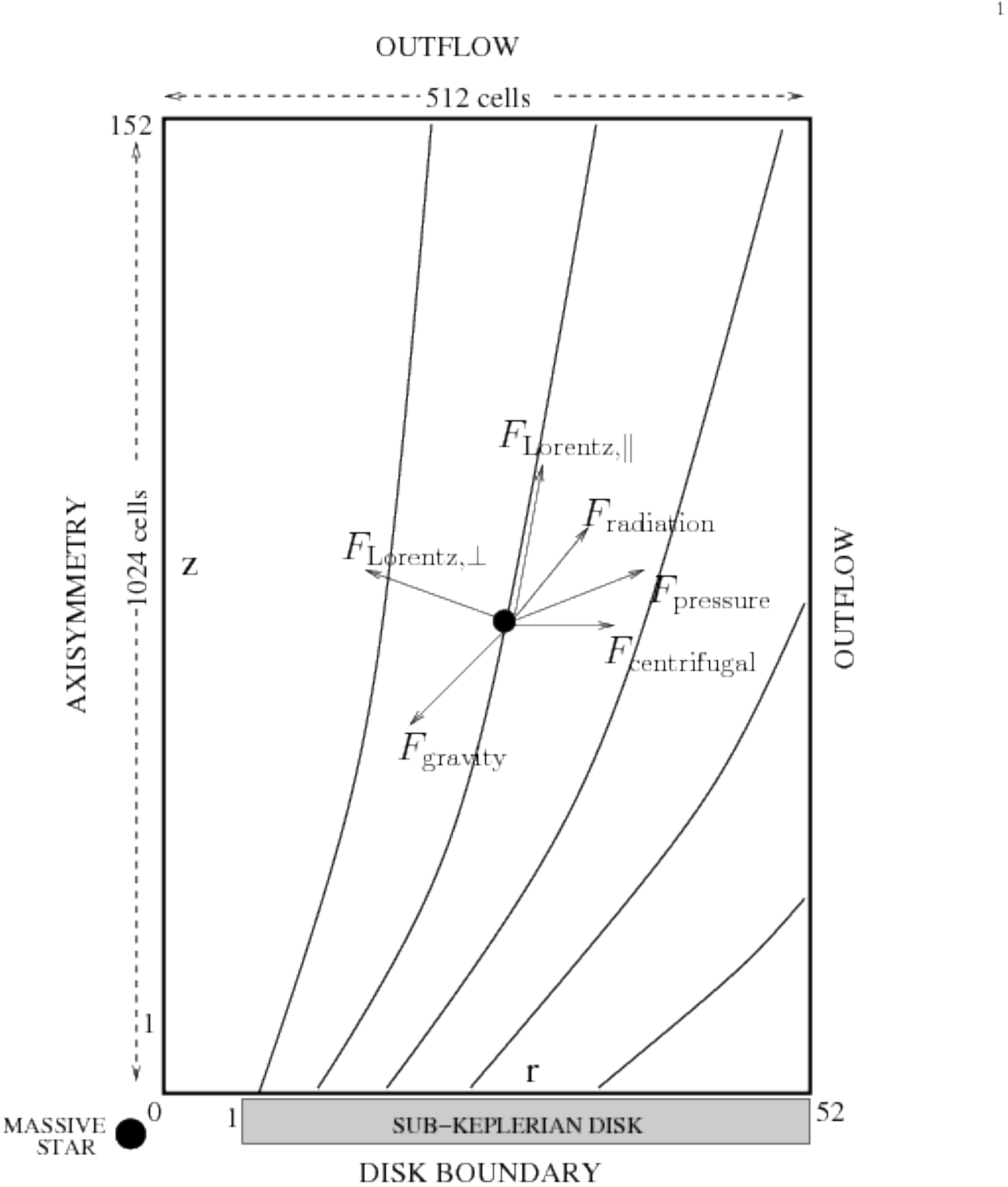}
\caption{Schematic view of the numerical setup along with definition of the boundaries conditions applied
for most of the simulation runs. 
The {\em lines} indicate poloidal magnetic field lines anchored in the underlying accretion disk 
which is in slightly sub-Keplerian rotation. 
The dark {\em black spot} represents a fluid element frozen onto that field line.
Vectors indicate and denote the various forces acting on the fluid element. 
See text for a detailed description of the different dynamically important forces.
}\label{numsetup}
\end{figure}

Figure~\ref{numsetup} displays the numerical setup used for our simulations. 
We also show the dynamically important forces that a fluid parcel experiences in an outflow from
a young massive star -
the gravitational force $\vec{F}_{\rm gravity}$ by the central star,
the Lorentz force components parallel $\vec{F}_{\rm Lorentz,\parallel}$ (accelerating) 
and perpendicular $\vec{F}_{\rm  Lorentz,\perp}$ (collimating),
the thermal pressure gradient $\vec{F}_{\rm pressure}$, and 
the centrifugal force $\vec{F}_{\rm centrifugal}$, which plays a vital role in particular for accelerating 
the flow from the disk surface. 
The essential point of this paper is that we also consider radiation
forces from the star and the disk ($\vec{F}_{\rm radiation}$, eq.~\eqref{radforce}). 
We consider the most dominant radiative source terms 
which are those due to the line driving mechanism from the central massive star and the 
underlying (hot) disk (see Fig~\ref{physetup}). 

\subsection{Initial conditions}
\label{ssec:Initcond}
We model the launching of the wind from an accretion disk representing the base of the outflow. 
The gravitational potential $\Phi$ is that of point star located at a slight offset 
$(-r_{\rm g},-z_{\rm g})$ from the origin to avoid singularity at r=z=0,
\begin{equation}\label{gravpot}
\Phi(r,z) \propto \left( (r+r_{\rm g})^{2} + (z+z_{\rm g})^{2} \right)^{-0.5},
\end{equation}
where $r_{\rm g} = z_{\rm g} = 0.21$ in all our simulations.

Initially we prescribe a hydrostatic equilibrium with a 
density distribution
\begin{equation}\label{initrho}
\rho(r,z) \propto \left((r+r_{\rm g})^{2} + (z+z_{\rm g})^{2}\right)^{-0.75}
\end{equation}
\citep{Ouyed:1997p634}.
The thermal pressure follows an polytropic equation of state $P = K \rho^{\gamma}$, 
with $\gamma = 5/3$. 
The magnitude of $K = (\gamma - 1)/\gamma $ is determined by the initial hydrostatic 
balance between gravity and pressure gradient. 
To begin with, all velocity components are vanishing. 
Such an initial setup is typical for jet launching
  simulations (e.g.\, \citealt{Ouyed:1997p634},
  \citealt{Krasnopolsky:1999p577}, \citealt{Fendt:2006p574}). 
The initial {\em{hot}} hydrostatic distribution of density and temperature does not play a
significant role in determining the flow structure as it is eventually
washed out by the {\em{cold}} [i.e.\, thermal
pressure unimportant] MHD flow that is launched from
the base.

The initial magnetic field is purely poloidal (in the $(r,z)$-plane) with a distribution derived from the potential
field $\vec{B}_{\rm p} = \nabla \times A_{\phi}$ with 
\begin{equation}
A_{\phi} = \frac{\sqrt{r^2 + (z + z_{\rm d})^2} - (z_{\rm d} + z)}{r}
\end{equation}
\citep{Ouyed:1997p634},
where $z_{\rm d}$ is considered as the dimensionless disk thickness.
Such a magnetic field configuration is force free.
Together with the hydrostatic density distribution this implies an initial
setup in force equilibrium.
The initial magnetic field strength is prescribed by choice of the plasma-beta $\beta_0$ 
at the inner launching point.

\subsection{Boundary conditions}
\label{ssec:boundary}
To choose the correct physical boundary conditions is of utmost importance for numerical 
simulations as they describe the astrophysical system under consideration.
In the present setup, we have to deal with four boundary regions (see Fig.~\ref{numsetup}).

\subsubsection{Axial boundary}
Along the jet axis an axisymmetric boundary condition is applied.
The normal and the toroidal components of vector fields 
(i.e. $B_{\rm n},B_{\phi},v_{\rm n},v_{\phi}$) change sign across the boundary, 
whereas the axial components are continuous. 
The density and the pressure are copied into the ghost zones from the domain.

\subsubsection{Equatorial Plane boundary}
The "disk boundary" is divided in two regions - the inner gap region extending from the
axis to the inner launching radius, $r < l_0$, and further out the disk region,
$r \ge l_0$, from where the outflow is launched.
 
The setup of this boundary is the most crucial for the outflow simulation. 
This is an "inflow" boundary with the boundary values determining the inflow of
gas and magnetic flux from the disk surface into the outflow.
Special care has to be taken to consider the causal interaction between the 
gas flow in the domain and in the ghost cells.

In order to have a causally consistent boundary condition (\citealt{Bogovalov:1997p680}, \citealt{Krasnopolsky:1999p577}, \citealt{Porth:2010p1041}), 
we impose the four physical quantities $\rho, \Omega^{\rm F}, P, E_{\phi}$. 
The toroidal electric field vanishes, $E_{\phi} = (\vec{v} \times \vec{B})_{\phi}=0$, as
result of the ideal MHD approximation. 
Thus, poloidal velocity and the poloidal magnetic field are parallel at the boundary, 
$\vec{v}_{\rm p} || \vec{B}_{\rm p}$. 
The angular velocity of the field line (Ferraro's iso-rotation parameter) $\Omega^{\rm F}$, 
which is one of the conserved quantities along the field line is fixed in time along the
boundary. 
We have chosen a Keplerian profile along the disk $\Omega^{\rm F}(r) \propto r^{-1.5}$. 
The poloidal velocity at the boundary is "floating", i.e. copied from the first grid cell 
into the ghost cell each time step. 
Thus, the mass flux at the disk boundary is not fixed but consistently derived by causal
interaction between the outflowing gas and the boundary value.

The inner gap region is prescribed as hydrostatic pressure distribution.
However, gas pressure gradient, gravity, and
centrifugal acceleration is considered for the radial force-balance in the disk,
leading to a sub-Keplerian rotation
$$\frac{v_{\phi}(l_0,0)}{v_{\rm{kep}}} = \sqrt{\chi}$$ 
with $\chi < 1$. 
Solving for the radial balance delivers the density profile along the boundary 
$$\rho_{\rm disk}(r,z) = \left(\frac{1}{1-\chi}\right) \rho(r,z).$$
Therefore, there is a density contrast between disk and initial corona (the domain).  
In order to avoid numerical problems at the interface of between gap and "inner disk" 
(ie. the inner jet launching radius), 
we smoothen this transition using the Fermi function, considering 
$$\chi = \left(\frac{\chi_0}{1 + \exp(-10(r-1))}\right). $$
The Fermi function is resolved by 16 grid cells.
The pressure distribution along the boundary is fixed
so as to have a cool disk which is three times denser than the initial
corona.  Thus, the wind emerging from such a
cool disk has a similarly cool temperature (lower than the
initial corona). 
Quantitatively, the average initial coronal temperature within the inner jet launching
area is around $5\times10^{5}$\,K, while the temperature of
the disk surface (i.e., the
boundary) in the same area reduces
to $5\times10^{4}$\,K at 10\,AU.

The magnetic field in the boundary is allowed to evolve in time 
to avoid any current sheets along the interface. 
When the simulation starts, the rotating disk winds-up the poloidal field and induces a
toroidal field component.
Constraining the field line angular velocity $\Omega^{\rm F}$ to be constant in time, 
we need to adjust the rotational velocity of the gas in the boundary,
\begin{equation}\label{fixomegaf}
 v_{\phi} = r\Omega^{\rm F} + \frac{\eta}{\rho} B_{\phi},
\end{equation}
where the mass load of a field line $\eta = (\rho v_{\rm p})/B_{\rm p}$ 
is again a conserved quantity in stationary MHD.

\subsubsection{Outflow boundaries}
The right and top boundaries (see Fig.~\ref{numsetup}) are defined as outflow boundaries.
The canonical outflow conditions (zero gradient across the boundary) are imposed to all 
scalar quantities and vector components, except for the magnetic field.
The toroidal magnetic field component $B_{\phi}$ in the boundary is obtained by requiring 
the gradient of the total electric current $I$ to vanish, 
whereas the poloidal field components are estimated by setting the toroidal current 
density $j_{\phi}$ to zero \citep{Porth:2010p1041}.
Having this new outflow condition implemented in the standard PLUTO code, 
we ensure that there is no artificial collimation due to boundary effects.

%
%
\section{Parameter survey}
\label{sec:parametersurvey}
\subsection{Choice of governing parameters}
\label{ssec:choicepara}
\begin{table}
\begin{center}
\caption{Summary of the dimensionless parameters to study the impact of 
radiation forces on the outflow dynamics}
\begin{tabular}{ c  c }
\hline\hline
\noalign{\smallskip}
Parameter & Definition \\
\hline
\noalign{\smallskip}
Eddington ratio   & \multirow{2}{*}{$\Gamma_{\rm e} \equiv$ \large{$\frac{\sigma_{e}L_{*}}{4\pi c GM_{*}}$}} \\
(proxy for stellar luminosity) &                                   \\
&\\
\noalign{\smallskip}
\hline
\noalign{\smallskip}
Luminosity ratio & \multirow{2}{*}{$\mu \equiv$ \large{$\frac{L_{\rm acc}}{L_{*}} = \frac{GM_{*}\dot{M}_{\rm{acc}}}{R_{*}L_{*}}$}} \\
(proxy for disk luminosity)&             \\
&\\
\noalign{\smallskip}
\hline
\noalign{\smallskip}
Radius ratio &\multirow{2}{*}{$\Lambda \equiv$ \large{$ \frac{R_{*}}{l_0}$}} \\
(proxy for stellar radius) &  \\
&\\
\noalign{\smallskip}
 \hline
\noalign{\smallskip}
Initial plasma beta & \multirow{2}{*}{$\beta_0 \equiv$\large{$ \frac{8\pi P_0}{B_0^{2}}$}} \\
(proxy for magnetic field  strength)&  \\
&\\
\noalign{\smallskip}
 \hline
\noalign{\smallskip}
Line-force parameters $Q_0$ and $\alpha$ &prescribes $M(\mathcal{T})$
using Eq.~\eqref{lfGayley} \\
&\\
\noalign{\smallskip}
\hline
\end{tabular}\label{ParaSummary}
\end{center}
\end{table}

The three main parameters we apply for a comprehensive study are the 
(i) central stellar mass $M_{*}$, 
(ii) plasma-beta $\beta_0$, and 
(iii) the line force parameter $\alpha$
(see Table~\ref{ParaSummary}).
We also find that the magnitude of the density at the launching base of the flow 
$\rho_0$, which is a free parameter in our simulations, strongly affects the 
flow characteristics (see discussion in \S~\ref{sec:rhobase}).

\begin{figure}
   \centering
 \includegraphics[width=1.0\columnwidth]{\figurepath/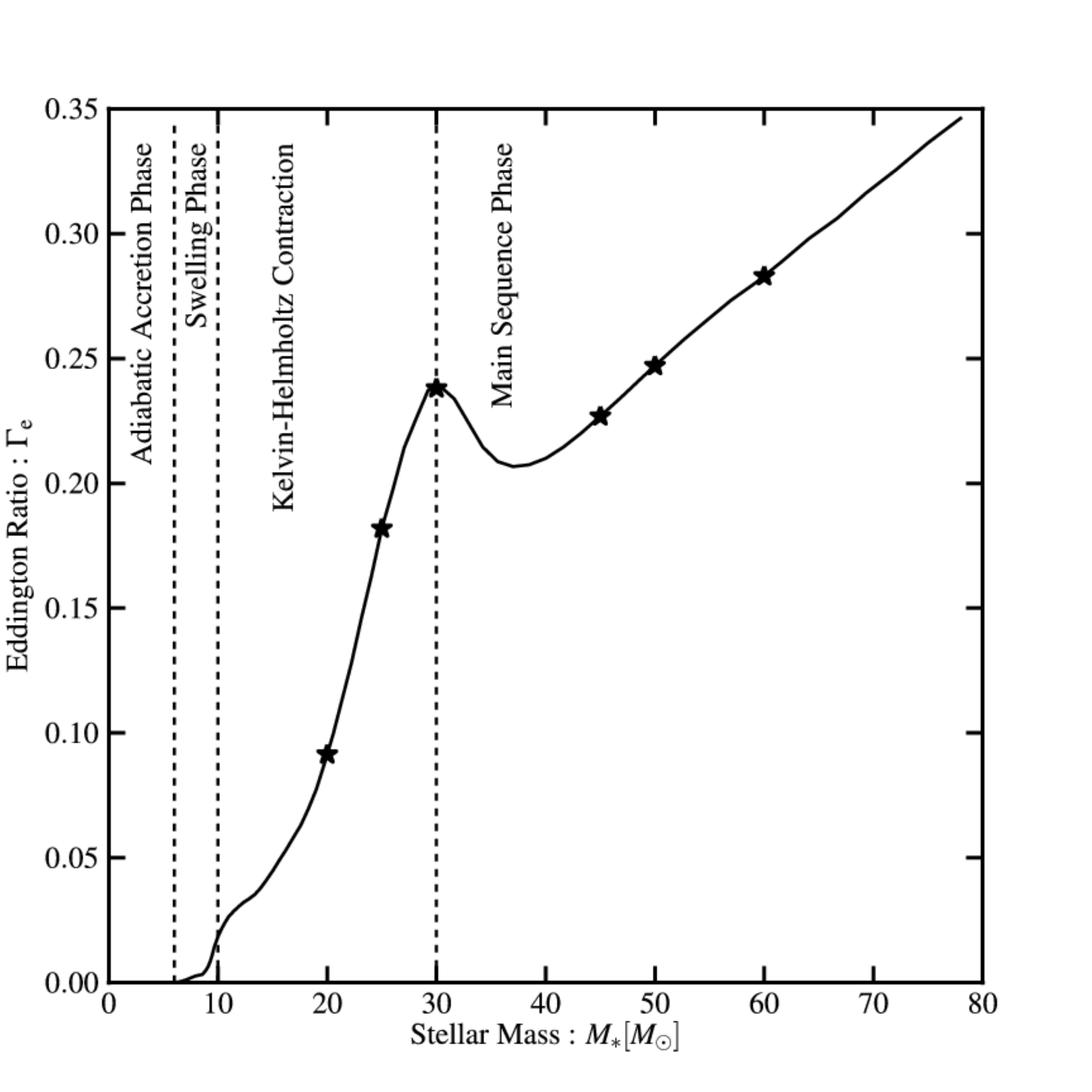}
   \caption{Variation of the dimensionless parameter $\Gamma_{\rm e}$ with stellar 
   mass $M_{*}$ from \cite{Hosokawa:2009p4005}.
   The  \textit{solid black line} indicates the stellar mass evolution in the 
   Hosokawa model.
   \textit{Stars} represent the Eddington ratio $\Gamma_{\rm e}$ for different 
   stellar masses in our parameter survey \S~\ref{sssec:masssurvey}. 
   The \textit{vertical dashed} lines mark different evolutionary stages for 
   massive young star accreting rapidly at the rate of $10^{-3}\,\msun\,\rm yr^{-1}$. 
   The curve is obtained using tabulated evolutionary parameters kindly provided 
   by Takashi Hosokawa (private communication).}
   \label{hosogammae}
\end{figure}

We have performed a large parameter study with respect to the central
star. The stellar mass considered in our model ranges from
20\,$\msun$ to 60\,$\msun$. 
The size and the luminosity of these stars are obtained from the literature
stellar evolution models (\citealt{Hosokawa:2009p4005}). 
The dimensionless parameter that controls the luminosity of the central star 
is  $\Gamma_{\rm e}$ (see Table~\ref{ParaSummary}). Its variation
with the stellar masses considered in our model is shown in Fig.~\ref{hosogammae}.
(Also see \S~\ref{sssec:masssurvey})

We have carried out simulations for three different values of 
magnetic field strength $\beta_0 = 1.0, 3.0, 5.0$. 
We fix the density at the base of the flow to the same value as for the simulations for 
different stellar mass.
Thus, for $30\,\msun$ star and an inner jet launching radius of 1\,AU, the different $\beta_0$ values 
correspond to a poloidal magnetic field of 11.5\,G, 6.6\,G, or 5.1\,G
at the inner launching point, respectively. 
The detailed description of the different simulations is shown in
Table ~\ref{VaryB}. 
The jet is launched by pure MHD processes. As it achieves a steady state after say $N_{\rm{MHD}}$ inner rotations,
we switch on the radiative line forces by the central star and run the simulation
for another $N_{\rm{RAD}}$ inner rotations. 

The stellar line force in our model is quantified by the parameters
$Q_0$ and $\alpha$. In principle, these parameters depend on the degree
of ionization in the flow, however, for simplicity, we neglect this dependence
and assume that their values remain the same for a particular simulation run. 
The line force is independent of $Q_0$, however, the its dependence on $\alpha$ 
is strong \cite{Gayley:1995p1968}.
We fix the line strength parameter $Q_0= 1400$ for all runs. 
We carry out simulations with three values of $\alpha = 0.55, 0.60,
0.65$. Although the differences in $\alpha$ seems to be small, they lead to considerable variation 
in the magnitude of the line forces \citep{Gayley:1995p1968}. These
values of $\alpha$ and $Q_0$ are consistent with the empirical
values obtained for evolved massive stars
(\citealt{Abbott:1982p900},\citealt{Gayley:1995p1968}). Since, we have no empirical values for these
parameters during the formation stage, the present
model uses similar values of line force parameters as
obtained from evolved massive stars.

\subsection{Quantifying the degree of collimation}
\label{ssec:quantcoll}

There are in principal several options to quantify the collimation degree of jets. 
\cite{Fendt:2002p1135} suggested to compare the mass fluxes in radial and vertical direction 
of the jet as a measure of the jet collimation. 
However, in general the choice of a "floating" inflow  boundary condition and the
subsequent time-dependent change of mass fluxes makes it difficult to
use them as a measure for the degree of collimation.
Therefore, in this work, we quantify the jet collimation by the opening angle $\phi$ of 
the field lines (which are equivalent to velocity streamlines in a steady state). 
For comparison, we measure the opening angle along a certain field line at two critical points along 
that field line viz. the Alfv$\acute{e}$n point $\phi_{\rm A}$ and
the fast magneto-sonic point $\phi_{\rm F}$. Note that this is not the
asymptotic jet opening angle.

In order to estimate the amount of de-collimation $\Delta \zeta[\%]$ by radiation
forces, we measure the angular separation of field lines resulting from $N_{\rm{MHD}}$ disk rotations 
from those with after $N_{\rm{RAD}}$ disk rotations,
\begin{equation}
\label{delzeta}
\Delta\zeta[\%](s) = 100 \left(\frac{\phi(s,N_{\rm RAD})-\phi(s,N_{\rm MHD})}{\phi(s,N_{\rm MHD})}\right),
\end{equation}
where $s$ measures of path along the field line. 
A positive value of $\Delta \zeta[\%]$ implies de-collimation whereas negative values would imply collimation of the jet by radiative forces.

\begin{table*}
\begin{center}
\caption{Parameter study of stellar radiation line-force effects on MHD disk jets for different 
  field strength and different stellar mass. 
  We apply the same physical density at the jet base $\rho_0 = 5.0\times10^{-14} \rm{g\,cm}^{-3}$, 
  the same inner launching radius $l_0 = 1\,\rm{AU}$, and 
  the same line-force parameter $Q_0=1400$ for all runs. 
  The simulations are performed for $N_{\rm{MHD}} = 319$ inner rotations in pure MHD, followed
  by and $N_{\rm{RAD}} = 319$ rotations, with switched-on radiative forces. 
  A measure for the jet opening angle is given at the MHD critical points, $\phi_{\rm{A}}$, $\phi_{\rm{F}}$
  and along the field line rooted at $r = 5.0\, AU$. 
  $\Delta \zeta[\%]$ denotes the percentage difference of the opening angles between the steady-state flows
  for pure MHD and including radiation force along the field line rooted at the same radius. 
  The vertical mass flux $\dot{M}_{\rm vert} [\msun\,\rm{yr}^{-1}]$ is measured along the top cells in the 
  domain. 
  The first ten simulation runs in the table apply a ''floating'' boundary condition for the injection velocity, 
  while for last three runs a fixed-mass-flux boundary condition is applied (see \S~\ref{ssec:injectmflux}).
 }
\begin{tabular}{c c c c c c c  c }
\hline\hline
\noalign{\smallskip}
Run ID & $M_{*} [\msun]$ & $\beta_0$ &$\alpha$ & $\phi_{\rm{A}}$ & $\phi_{\rm{F}}$ & max[$\Delta \zeta[\%]$] & $\dot{M}_{\rm vert} [\msun\,\rm{yr}^{-1}]$ \\
\noalign{\smallskip}
\hline
\noalign{\smallskip}
M30a055b1 & 30 & 1.0 & 0.55 & 21.98 & 16.45 &5.0&$3.1\times10^{-5}$\\
M30a055b3 & 30 & 3.0 & 0.55 &25.63 &20.68   &17.0&$2.9\times10^{-5}$\\
M30a055b5 & 30 & 5.0 & 0.55 & 26.05 & 23.40 &34.5&$2.5\times10^{-5}$\\
M30a055b5-c & 30 & 5.0 & 0.55 & 25.55 & 23.01 &33.9  &$2.4\times10^{-5}$\\
\noalign{\smallskip}
\hline
\noalign{\smallskip}
M20a055b5 & 20 & 5.0 & 0.55 & 20.44 & 15.57 &5.8&$1.6\times10^{-5}$\\
M25a055b5 & 25 & 5.0 & 0.55 & 23.27 & 19.94 &21.9&$2.4\times10^{-5}$\\
M45a055b5 & 45 & 5.0 & 0.55 & 26.84 & 24.35 &37.7&$3.0\times10^{-5}$\\
M50a055b5 & 50 & 5.0 & 0.55 & 28.86 & 26.03 &42.0&$3.3\times10^{-5}$\\
M60a055b5 & 60 & 5.0 & 0.55 & 31.45 & 29.25 &56.0&$3.7\times10^{-5}$\\
\noalign{\smallskip}
\hline
\noalign{\smallskip}
M60a060b5 & 60 & 5.0 & 0.60 & 23.73 & 21.48 &28.7&$3.2\times10^{-5}$\\
M60a065b5 & 60 & 5.0 & 0.65 & 20.93 & 17.28 &11.7&$2.9\times10^{-5}$\\
\noalign{\smallskip}
\hline
\noalign{\smallskip}
M25a055b3inj & 25 & 3.0 & 0.55 & 45.84 & 33.87 & 30.2 &$1.3\times10^{-5}$ \\
M30a055b3inj & 30 & 3.0 & 0.55 & 47.71 & 36.97 & 37.2 &$1.4\times10^{-5}$ \\
M50a055b3inj & 50 & 3.0 & 0.55 & 48.76 & 39.22 & 42.1 &$1.8\times10^{-5}$ \\
\noalign{\smallskip}
\hline

\hline
\end{tabular}\label{VaryB}
\end{center}
\end{table*}

\subsection{The outflow density at the disk surface - $\rho_0$ }
\label{sec:rhobase}

The radiative force term $F^{\rm{rad}}_{\rm{c}}$ is a 
product of the force multiplier and the continuum force. 
The force multiplier (Eq.~\ref{lfGayley}) depends on the physical
scalings $v_0$, $l_0$ and $\rho_0$
(see Eq.~\ref{nondimforcemult}). This does imply that each simulation
run is unique to the chosen set of scaling parameters
for a particular type of star.

In case of massive stars these values are currently difficult to measure very close to the
star (see \S~\ref{sec:modelsetup}). 
However, observationally derived values of mass flow rates in molecular outflows and jets 
around massive stars (\citealt{Beuther:2002p1149}, \citealt{Zhang:2005p4269}, \citealt{LopezSepulcre:2009p4143})
can be used to constrain the densities with prior assumption of inner
jet velocity. Typically measured values of the mass outflow rate are of the
order of $10^{-3}$ to $10^{-5}\,\msun\,\rm{yr}^{-1}$.

The density $\rho_0$ at the base of the flow ($z \sim 0$) can be estimated
by the mass flux launched from the base per unit time,
\begin{equation}
\dot{\rm M}_{\rm out} = 2 \pi \rho_0 v_0 l_0^2 \int_{r_{\rm c,min}}^{r_{\rm
    c,max}} r_{\rm c}^{1/2 - q} d r_{\rm c},
\label{mfluxobs}
\end{equation}
where the density at the base of the flow is $\rho(\rm r_{\rm
    c},z_{\rm c}=0) = \rho_0 r_{\rm c}^{-q}$. From
Eq.~\eqref{initrho}, q = 3/2.  We assume the matter is
launched from the disk surface with Keplerian speed (i.e. $v_{\rm
  z}(r_{\rm c},z_{\rm c}=0) = v_0 r_{\rm c}^{-1/2}$). 
The physical scaling for the density,
velocity and lengths are $\rho_0$, $v_0$ and $l_0$ respectively,
while $r_{\rm c}$ is the non-dimensional length unit.
Thus, the density scaling $\rho_0$ can be expressed as
\begin{equation}
\label{valrhobase}
\rho_0 = \frac{\dot{\rm M}_{\rm out}}{2 \pi v_0 l_0^2}\left[\rm{ln} \left(\frac{r_{\rm
    c,max}}{r_{\rm c,min}}\right)\right]^{-1}.
\end{equation}
Using typical observed mass outflow rates for young massive stellar
jets, we calculate $\rho_0 \sim 10^{-13} - 10^{-15}\,\rm{g}\, \rm{cm}^{-3}$.

The physical value of the density at the base of the flow is in the denominator of Eq.~\ref{nondimforcemult}, 
and therefore affects the magnitude of the line-driving force significantly.
Increasing the density by, say, a factor of 10, decreases the line force at least 
by a factor of $10^{\alpha}$. 
For the values of $\alpha$ considered her, this magnification could be as large as 3. 
Such a change in the radiative force has clearly a notable impact on the outflow dynamics,
in particular on its degree of collimation. 
The description of our simulations with different base density is given in Table~\ref{Varyrhobase}.
\begin{table*}
\begin{center}
\caption{Parameter study of stellar radiation line-force effects on MHD disk jets for different 
  jet density at the base of the flow with a stellar mass of $30\,\msun$.
  For all runs, plasma-$\beta_0 = 5$, $l_0$ = 1\,AU and line-force parameters $Q_0=1400$ and $\alpha$ = 0.55.
 }
\begin{tabular}{c c c c c c}
\hline\hline
\noalign{\smallskip}
Runs & $\rho_0 [\rm{g}\,\rm{cm}^{-3}]$ & $\phi_{\rm{A}}$ & $\phi_{\rm{F}}$
& max[$\Delta \zeta[\%]$] &$\dot{M}_{\rm vert} [\msun\,\rm yr^{-1}]$\\
\noalign{\smallskip}
\hline
\noalign{\smallskip}
M30a055b5d1 & $3.0\times10^{-14}$ & 30.30 & 27.79 &47.7&$1.5\times10^{-5}$\\
M30a055b5 & $5.0\times10^{-14} $& 26.02 & 23.36 &34.5&$2.4\times10^{-5}$ \\
M30a055b5d2 & $1.0\times10^{-13} $& 22.91 & 19.47 &19.6&$4.3\times10^{-5}$\\
M30a055b5d3 & $5.0\times10^{-13}$ & 20.42 & 15.03 &4.0&$1.9\times10^{-4}$ \\
\noalign{\smallskip}
\hline
\end{tabular}\label{Varyrhobase}
\end{center}
\end{table*}

\begin{table*}
\begin{center}
\caption{Parameter study of disk radiation line-force effects on MHD disk jets.
  All simulation runs apply the same stellar parameters and line-force 
  parameters as M30a055b5, however a different inner launching radius of only 0.1\,AU. 
  Additionally, two more dimensionless parameters for the disk radiation force 
  $\mu = 0.4644$ and $\Lambda =0.4969$ are prescribed (see Table~\ref{ParaSummary}).}
\begin{tabular}{c c c c c}
\hline\hline
\noalign{\smallskip}
Run & $\rho_0 [\rm g\,\rm{cm}^{-3}$] & max[$\Delta \zeta[\%]$]
&$\dot{M}_{\rm vert} [\msun\,\rm{yr}^{-1}]$&
Comments\\
\noalign{\smallskip}
\hline
\noalign{\smallskip}
Disk1 & $5.0\times10^{-14} $ & -1.8 & $6.1\times10^{-7}$  & reaches a steady state\\
Disk2 & $5.0\times10^{-15} $ & 3.1 & $7.9\times10^{-8}$  & remains unsteady\\
\noalign{\smallskip}
\hline
\end{tabular}\label{Diskandinj}
\end{center}
\end{table*}

%
%
\section{Results and discussion}
\label{sec:results}
The radiation field in massive star forming regions may play a crucial role in modifying the 
dynamics of outflows and jets. 
In order to disentangle the effects of radiative forces from the pure MHD jet
formation, we decided to follow a two-step approach.
We first i) launch a pure MHD disk jet and wait until it has reach a steady state
(after about $N_{\rm MHD} = 320$ rotations).
We then ii) switch on the radiation line-forces and allow the jet flow to further develop 
into a new dynamic state. In some cases a new steady-state can be reached, in other
cases an unsteady solution develops.

For the radiation forces we consider line-forces exerted by
 i) the luminous massive young star, and those by
ii) the surrounding accretion disk. 

\subsection{Jet de-collimation by the stellar radiation field}
The radiation line-force from the central point star is determined by
applying Eq.~\eqref{flinestar}. 
For a detailed physical analysis of the effects of line-forces from the star alone, 
we have chosen simulation M30a055b5 as reference simulation run (Table~\ref{VaryB}).
The reference run is parametrized for a stellar mass of $30\msun$, 
surrounded by an accretion disk with an inner jet launching radius of 1\,AU, 
and a density of $5.0\times10^{-14} {\rm g\,cm^{-3}}$ at this radius. 
The initial poloidal magnetic field strength is fixed by a
plasma-$\beta_0 = 5.0$.
The radiative forces from the star are defined by the
line force parameters $Q_0 = 1400.0$ and $\alpha =0.55$.

\begin{figure}
\centering
\includegraphics[width=1.0\columnwidth]{\figurepath/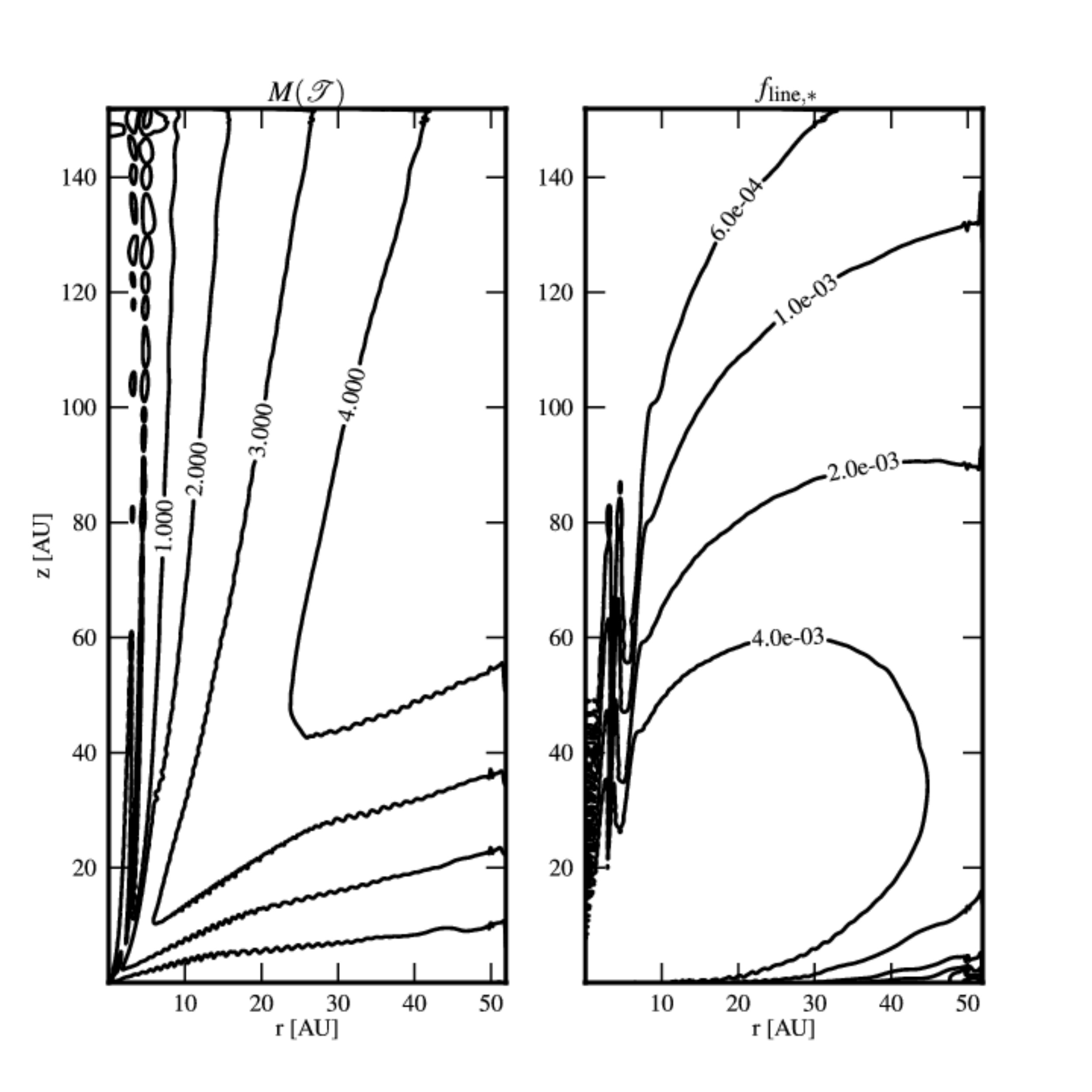}
\caption{Contours of force multiplier $\rm{M}(\mathcal{T})$ and the specific line radiation force 
of a central star [in physical units] for the reference simulation M30a055b5 after 
$N_{\rm{rot}} = 638$ inner rotations.}
\label{lineforce_img}
\end{figure}

The spatial distribution of the force multiplier $M(\mathcal{T})$ as well as the 
specific stellar line force for the reference simulation are both shown in 
Fig.~\ref{lineforce_img}. 
The magnitude of the force multiplier peaks in the top-right low-density regions 
of the flow. 
This is expected as the force multiplier increases with decreasing density as 
shown in Eq.\ref{flinestar}. 
In order to calculate the true radiation force, the force multiplier must be 
convolved with the continuum radiation from the star, resulting in a line-force 
distribution with a maximum close to the central star
(see Fig.~\ref{lineforce_img}).

\begin{figure*}
\centering
\includegraphics[width=16cm]{\figurepath/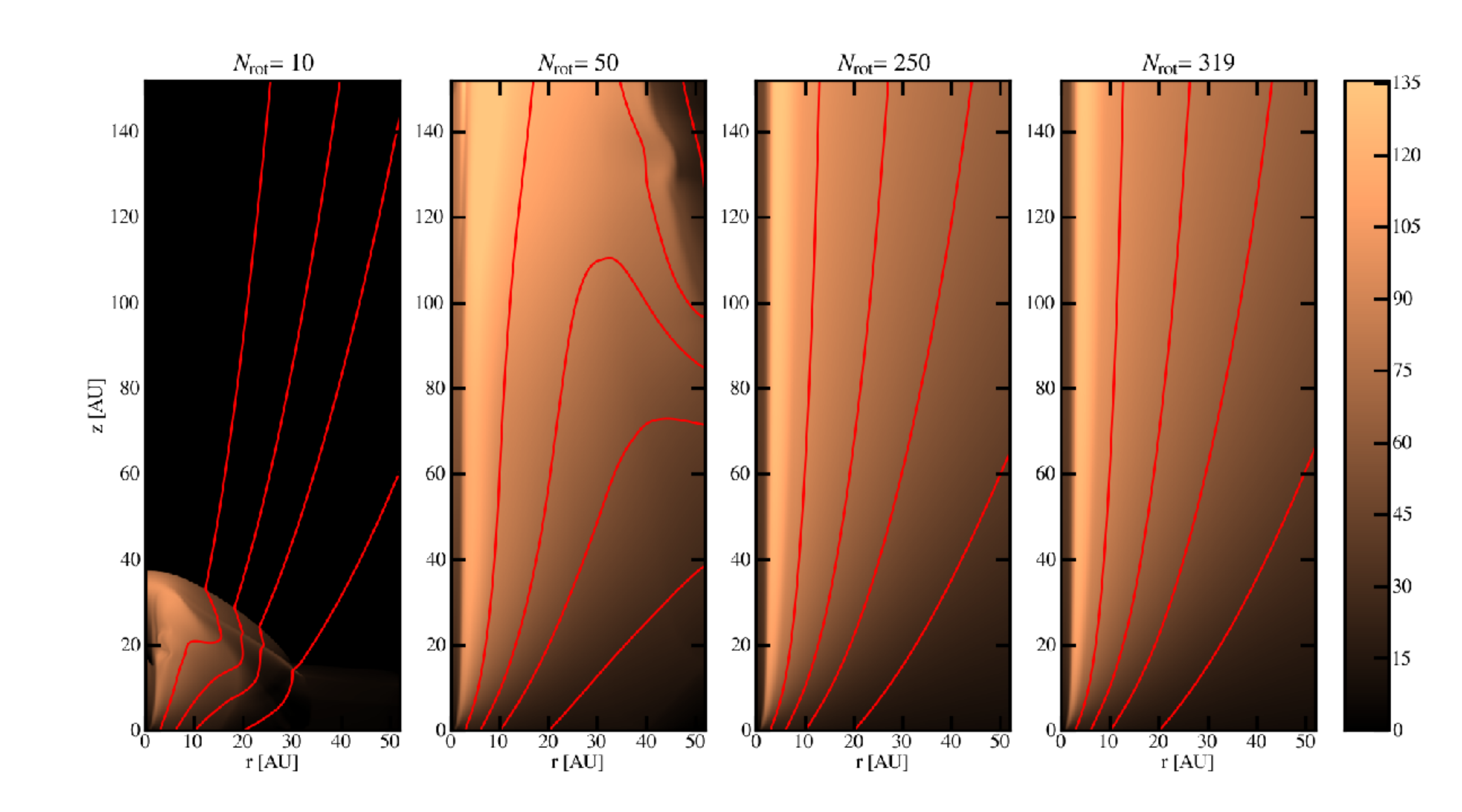}\\
\includegraphics[width=16cm]{\figurepath/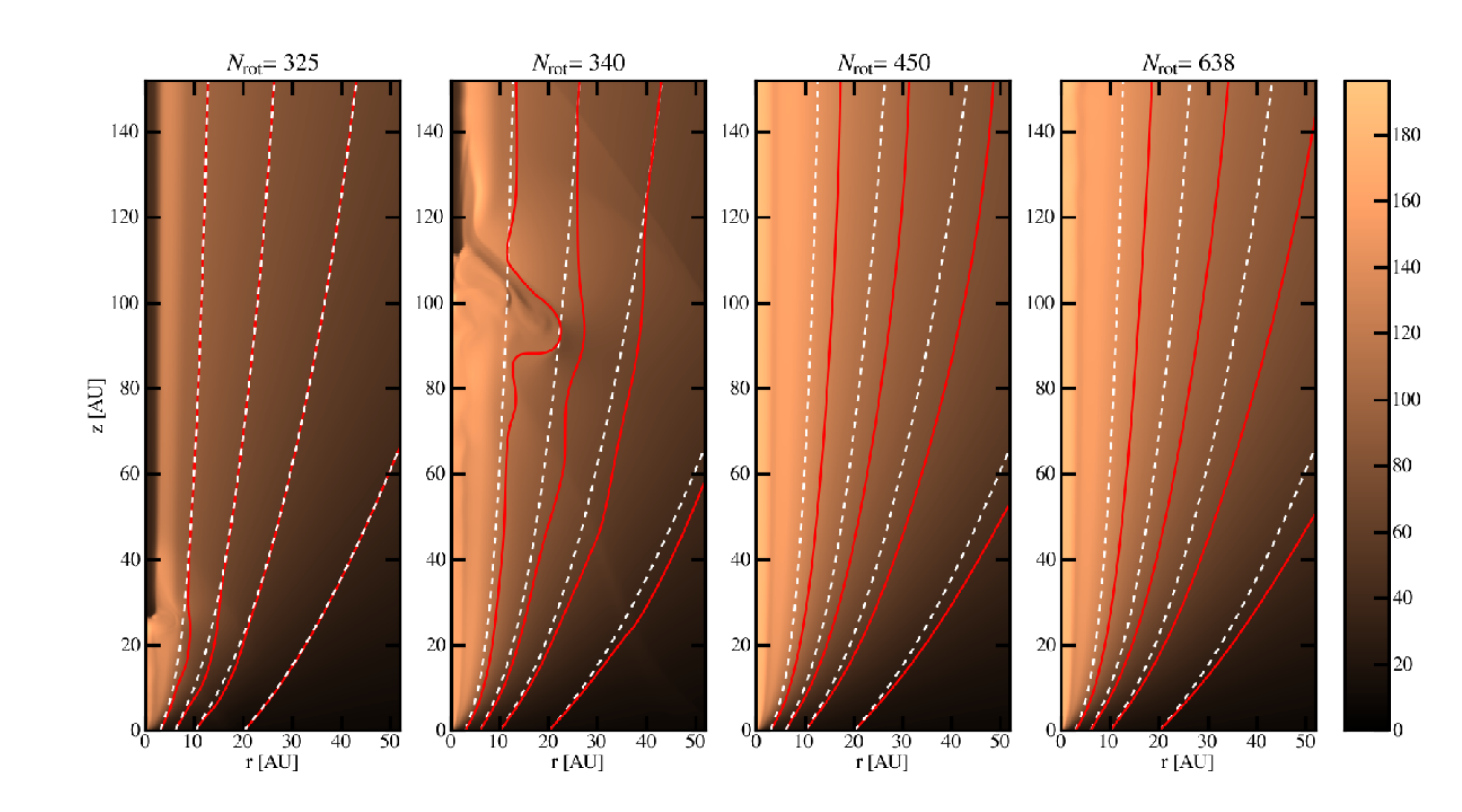}\\
\caption{Evolutionary sequence of the reference simulation M30a055b5. Shown is the jet vertical 
velocity distribution (color coding) and the poloidal magnetic field lines.  
The color bar represents the velocity scale in km\,s$^{-1}$ for each row of panels. 
The series of images are taken at subsequent inner rotations as mentioned on the top of each image. 
The \textit{solid red lines} are the poloidal magnetic field lines. 
To illustrate the impact of radiation forces, we also show the field lines obtained after the steady-state MHD 
flow for comparison as \textit{white dashed lines} in the lower four panels. 
}
\label{evolvevz}
\end{figure*}

The time evolution of the emerging jet\symbolfootnote[2]{Full movies are available for 
download at 
\url{http://www.mpia.de/homes/fendt/vaidya/research_bv.html}
} 
is shown in Fig.~\ref{evolvevz}, where we display the vertical jet velocity $v_z(r,z)$ and the 
poloidal magnetic field structure for reference simulation M30a055b5.
We clearly see the change from the pure MHD flow (\textit{top} panels)
to the situation when stellar radiation forces are considered (\textit{bottom} panels).

Material which is injected from the disk surface (\textit{bottom boundary}) 
is "frozen" on these field lines (ideal MHD).
Initially, the plasma is accelerated magneto-centrifugally and gains
substantial speed, producing a bow shock as it propagates.  
The bow shock leaves the domain after $\sim 60$ rotations.
Inertial forces of the outflowing mass flux induce a strong toroidal magnetic field
component resulting in magnetic hoop stresses which self-collimate the magnetic field 
structure together with the hydrodynamic mass flux. 
Eventually, when all the dynamical forces along the field line are balanced again,
the flow achieves a steady-state.
The steady state magnetic field configuration obtained
after the pure MHD flow is shown as \textit{white dashed} lines in Fig.~\ref{evolvevz}.

At this stage we "switch on" the line-forces of the central star. 
Immediately, the emergence of a fast axial flow is visible. 
This flow is unsteady, first forming a knotty structure, and then, when approaching the 
upper boundary, stabilizes to a steady axial flow. 
Jet de-collimation by stellar radiation forces is indicated by the fact that the 
\textit{red solid} field lines in the \textit{bottom} panels of Fig.~\ref{evolvevz} do 
open-up significantly as compared to field lines in steady-state pure MHD simulation. 

\begin{figure}
\centering
\includegraphics[width=1.0\columnwidth]{\figurepath/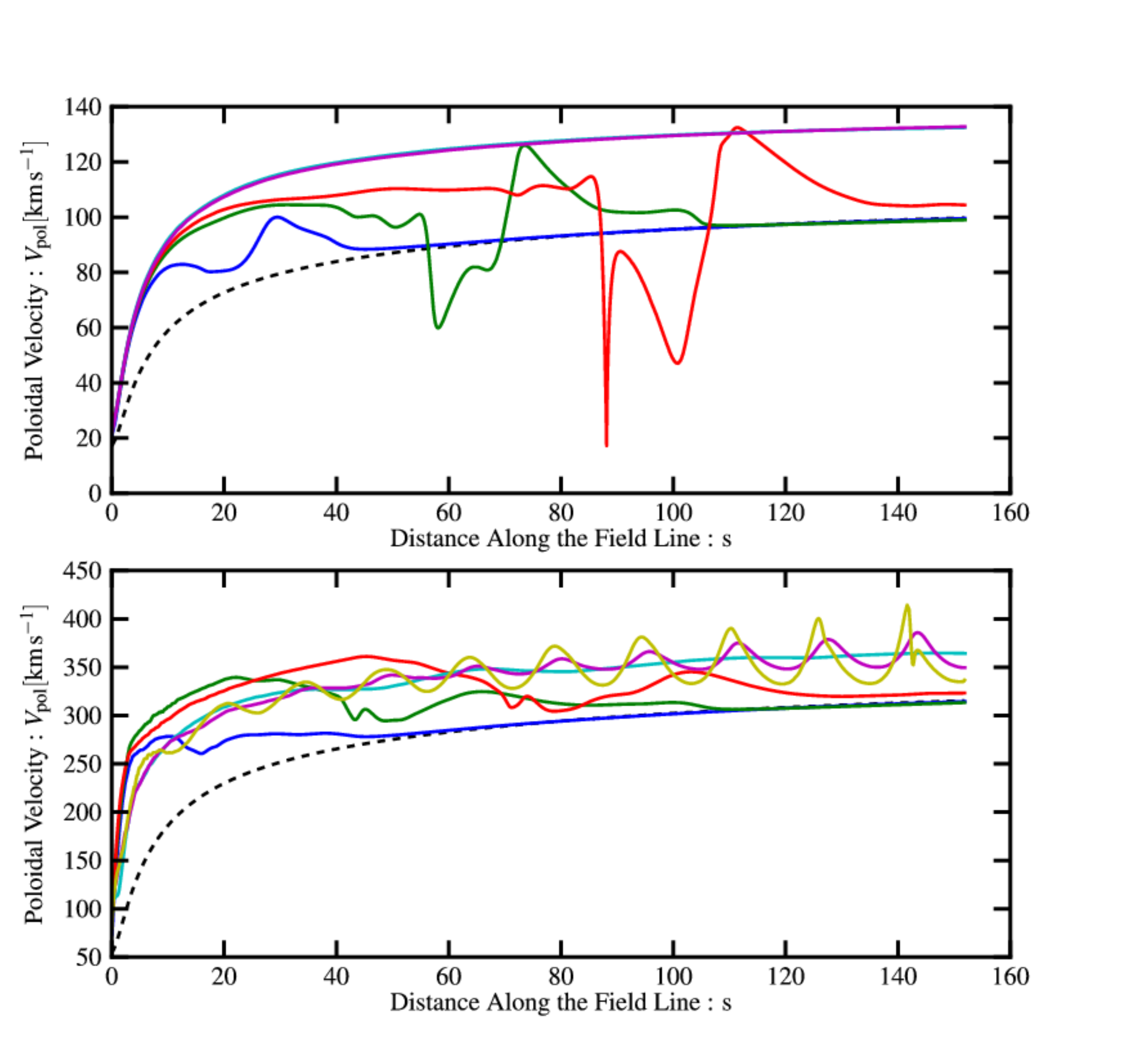}
\caption{Poloidal velocity along the field line rooted
  at r = 3\, AU for the two simulations
M30a055b5 \textit{top panel} and  Disk2 \textit{bottom panel}. The
distance along the field line $s$ is measured in AU.
The various colored \textit{solid} curves indicate the velocity
profile at different inner disk rotations $N_{\rm rot}$
(\textit{blue} for $N_{\rm rot}$= 325, \textit{green} for $N_{\rm rot}$= 333, \textit{red} for $N_{\rm rot}$=340,\
 \textit{cyan} for $N_{\rm rot}$= 510, and \textit{magenta} for
 $N_{\rm rot}$ = 638). 
The \textit{black dotted} line in both panels correspond to the last time step for the pure MHD flow. 
The \textit{solid yellow} line shown in the bottom panel indicates the poloidal velocity profile 
after 1500 inner rotations.}
\label{ldinstablefig}

\end{figure}

When the radiative force is switched on, a shock front begins to propagate into
the steady MHD jet (Fig.~\ref{evolvevz}).
This happens, because the additional radiation forces lead to an initial disturbance 
at the base of the flow, which is then propagated outwards. 
The effect of line-forces resulting in a series of propagating shocks is best seen in 
the poloidal velocity profile along a magnetic field line (see Fig.~\ref{ldinstablefig}).  
The series of shocks eventually propagate out of the domain, and the flow attains a steady 
state again.
However, the asymptotic outflow velocity which is then achieved is enhanced by a factor 
1.5 - 2 compared to the pure MHD flow. 

We have also carried out a run, M30a055b5-c, that
  includes the radiation force right from the beginning along with
the MHD flow. This flow evolves into the same configuration as our
reference run where radiation force is ''switched on'' later. 
This proves that the initial conditions do not
affect the final state of the flow. In our study, we prefer to use the
step-by-step approach as it is computationally less expensive.

In summary, we find that radiation forces modify the MHD disk jet essentially in terms
of collimation, but also in terms of acceleration and terminal speed.

\subsubsection{Analysis of the force-balance in the outflow }
\begin{figure*}
\centering
\includegraphics[width=16cm]{\figurepath/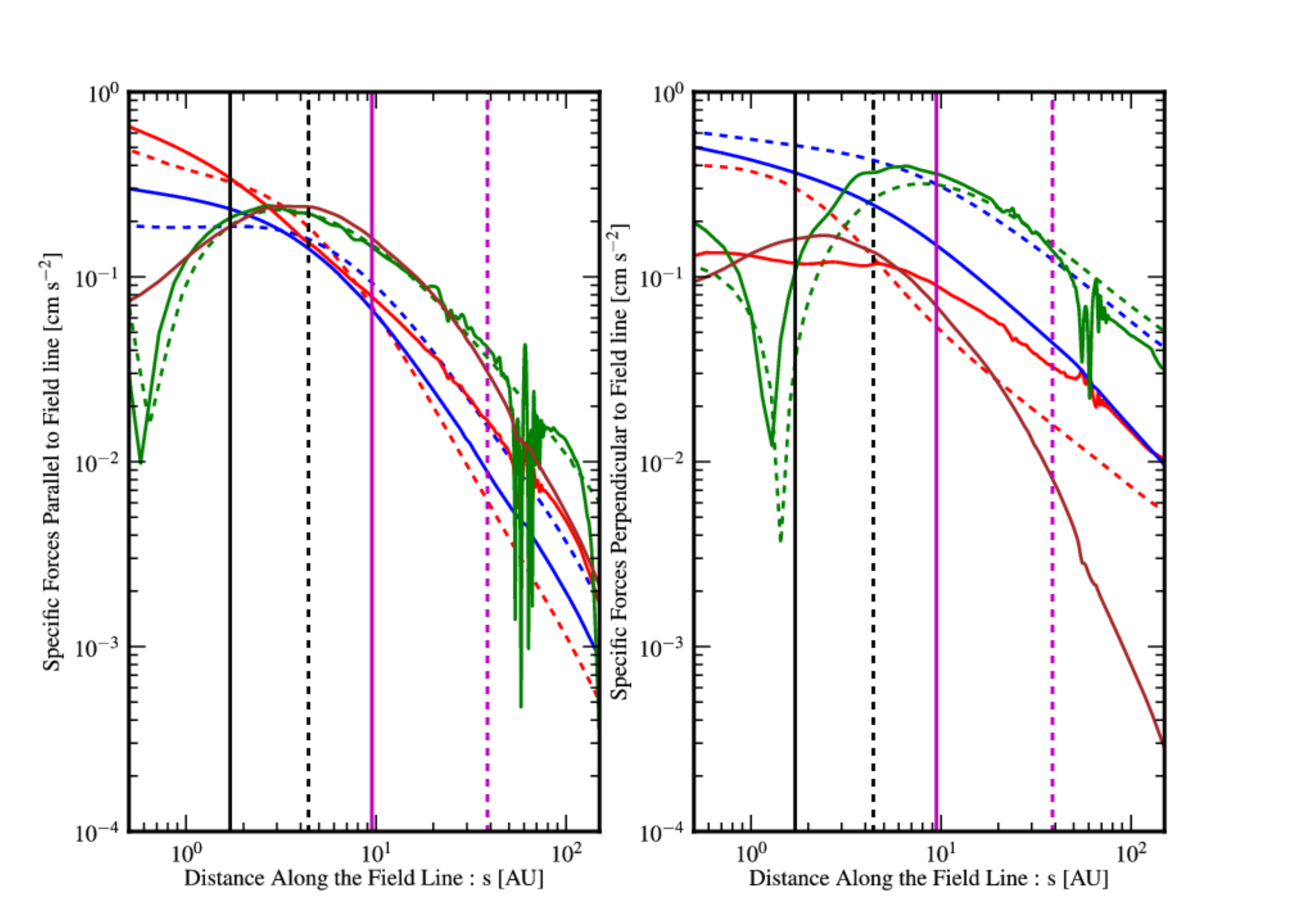}\\
\caption{Comparison of all specific force terms projected {\em parallel}[left] and {\em perpendicular}[right] to the field line 
rooted at r=5.0\,AU for the run M60a055b5. 
Colors specify different specific force terms in dyne cm$^{3}$ g$^{-1}$ - the Lorentz force (\textit{green}), the gas pressure gradient (\textit{red}), 
and the centrifugal force (\textit{blue}).
The \textit{dashed lines} corresponds to the respective force terms for the pure MHD run M60a055b5 (at steady-state), 
while the \textit{solid lines} represents the forces for the final time step of the simulation including radiative forces. 
The \textit{brown solid line} represents stellar radiation force term for the final time step. 
The \textit{black} and \textit{magenta} vertical \textit{solid lines} mark the position of the Alfv$\acute{e}$n and the 
fast magneto-sonic surface for the steady state flow including radiative forces, whereas the corresponding \textit{dashed lines}
are for the pure MHD flow.}
\label{forcecompfig}
\end{figure*}

Here we investigate the main forces affecting the outflow dynamics, 
comparing the magnitude of various force terms calculated at each grid point along
the field line rooted at r = 5.0\,AU for simulation run M60a055b5. 
Figure ~\ref{forcecompfig} shows such a comparison of specific forces projected {\em parallel} 
and {\em perpendicular} to the field line before and after considering line-forces due to 
stellar radiation.
\begin{figure}
\centering
\includegraphics[width=1.0\columnwidth]{\figurepath/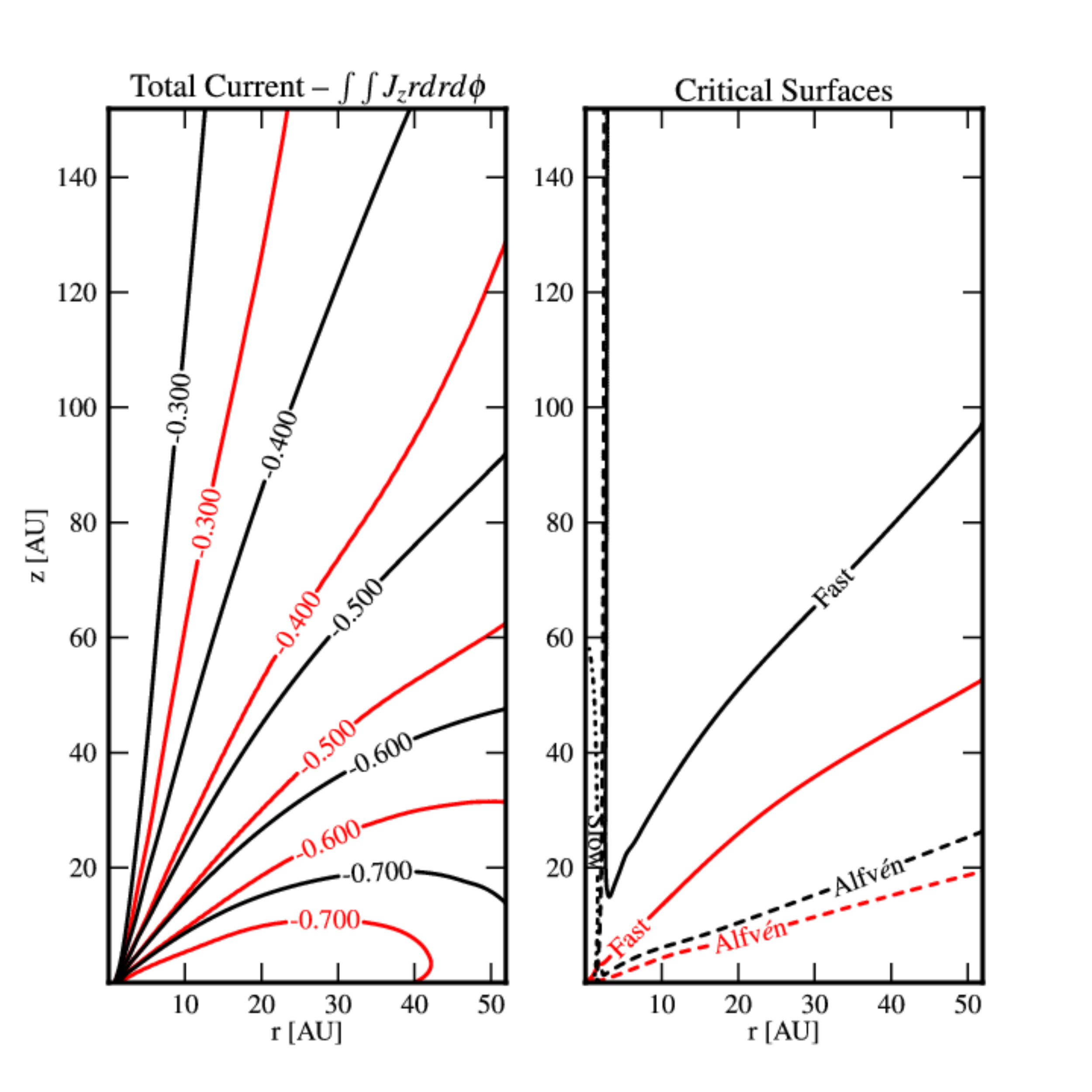}
\caption{Contours of the total poloidal electric current distribution
(\textit{left}), and the location of the MHD critical surfaces (\textit{right}). 
In both panels, the pure MHD flow is represented by \textit{black solid lines}, while the radiative MHD flow is shown 
with \textit{red solid lines}.}
\label{contocomp}
\end{figure}
With respect to acceleration, the most striking feature is the enhanced specific pressure gradient force. 
When radiative forces are considered, the steady MHD flow material at higher altitudes is adiabatically 
compressed from the underlying accelerated material leading into a shock that propagates out of the domain. 
The resultant flow has higher temperature or an increase in thermal pressure. 
We also find that the gradient of the thermal pressure higher above in the flow may increase substantially,
such that thermal pressure force and Lorentz force may become
comparable. In terms of collimation, it is evident from
Fig.~\ref{forcecompfig} that the profile of the perpendicular
component of the pressure gradient force along the field line becomes
flatter than in the pure MHD flow case.
Thus, the pressure force which was not important for pure MHD flows, now becomes a 
significant factor in governing the dynamics of the initial flow acceleration (disk wind).

Contrary, when the line-driving force is switched-on in the MHD jet, the centrifugal force 
becomes reduced by an order of magnitude particularly at high altitudes in the flow. 
In order to comprehend this variation of centrifugal force, it is useful to compare the  
conserved quantities of MHD, in particular the angular velocity of the field line 
$\Omega_{\rm F}$.
In the pure MHD simulation $\Omega_{\rm F}$ is conserved throughout the simulation.
This remains true also with radiative forces included, simply because $\Omega_{\rm F}$ is
fixed as boundary value (see \S~\ref{ssec:boundary}). 
Therefore, the azimuthal velocity at the base of the flow is given by Eq.~\eqref{fixomegaf}. 
At the same time, the additional line forces accelerate the flow (close to the base) to 
higher poloidal velocities. 
The magnetic flux at the base is fixed as a boundary condition.
Hence, the ratio of mass load to density $\eta/\rho$ in Eq.~\eqref{fixomegaf} increases, 
and since the toroidal magnetic field $B_{\phi}$ is negative, also the azimuthal velocity 
decreases. 
This eventually leads to a decrease of the specific centrifugal force when line-forces are
considered.
The decrease in centrifugal force further implies that the outflow rotates slower,
and that the acceleration close to the base is 
not only controlled by magneto-centrifugal forces, but also 
by the thermal pressure force and the radiation force (see Fig.~\ref{forcecompfig}).

Close to the base of the flow, Lorentz forces are dynamically not important.  
However, they peak at the Alfv$\acute{e}$n point of that particular field line. 
Beyond the Alfv$\acute{e}$n point, the Lorentz force becomes important and competes with other forces 
to govern the dynamics of the flow. When radiative force are
considered, the Lorentz force close to the 
base of the flow and also at higher altitudes do not show significant deviations from its values in the pure MHD case.  

Line-driven radiative forces have considerable impact on the poloidal electric current 
distribution and also on the critical MHD surfaces in the jet. 
The contour plots shown in Fig.~\ref{contocomp} compare the total poloidal current $I=rB_{\phi}=\int \vec{j}_{\rm p} d\vec{A}$ 
and position of critical surfaces before and after considering line-driven forces
in the reference simulation M30a055b5. 
The figure shows that by adding radiative forces (stellar luminosity), the corresponding 
poloidal electric current density contours are shifted closer to the base of the flow,
implying a lower toroidal field strength $B_{\phi}$ in these jets. 
We understand this results as a consequence of lack of jet rotation, thus less effective
induction of toroidal magnetic field.

Jet collimation in the conventional Blandford-Payne picture is caused by magnetic hoop stresses
(-$B_{\phi}^2/r$) of the azimutal magnetic field.
Whereas, the larger toroidal magnetic pressure gradient aids in de-collimating the outflow.  
A lower $B_{\phi}$ would not only weaken the hoop stress but also reduce the magnetic 
pressure ($\sim B_{\phi}^2$). 
The decrease of toroidal magnetic pressure would lead to a lower magnetic pressure gradient force. 
So, a balance arises between the hoop stress that collimates the flow
and the toroidal 
pressure gradient that de-collimates it. 
We observe that the field lines de-collimate to a wider configuration on addition 
of radiative forces, 
implying that the decrease of toroidal magnetic field has more impact on reducing 
of hoop stresses than reducing the magnetic pressure gradient, 
thus eventually de-collimating the flow.

Considering radiative forces also results in lowering the location of MHD critical surfaces 
(viz. slow magneto-sonic, fast magneto-sonic and Alfv$\acute{e}$n surface)
(see Fig.~\ref{contocomp}). 
The pure MHD flow is launched marginally super-slow
and sub-Alfv$\acute{e}$nic. 
The flow near the axis is, however, sub-slow, due to the boundary condition of 
conserving the initial hydrostatic density and pressure distribution in the gap
region between axis and disk.
The flow speed at the critical points depends on the magnetic flux and mass density 
in the flow.
These quantities remain approximately unaltered when radiative forces are considered,
thus the magneto-sonic wave speed remains similar as well. 
Since now the radiative line-forces accelerate the wind further to
higher velocities, the critical surfaces shift to lower
altitudes in the flow (i.e. close to the base of the flow).    

We finally note that radiative forces have both a direct as well as an indirect effect in
modifying the collimation properties of the flow. 
The radiation force from the star {\em directly} affects the flow dynamics at its base and 
close to the star simply by transfer of momentum (Eq.~\ref{momcons}), eventually leading 
to a flow de-collimation.
{\em Indirectly}, the radiation field also enhances the thermal pressure in the flow 
(Eq.~\ref{encons1}) such that the specific pressure force becomes comparable to the Lorentz 
force - further de-collimating the flow.
The resulting opening angles as measured from our reference run at various critical points 
along the field lines are listed in Table ~\ref{VaryB}. 
For M30a055b5, the final state of the jet has a steady mass outflow rate of
$2.5\times10^{-5} \msun {\rm yr^{-1}}$. 
The opening angles are $26^\circ$ and $23^\circ$ at Alfv$\acute{e}$n and fast point
respectively for the field line which has a foot point r = 5\, AU. 
The maximum percentage separation $\Delta\zeta[\%]$ between the opening angles of this 
field line in pure MHD flow and that for the flow with radiative forces is 34.5\%. 
A significantly positive value of $\Delta\zeta[\%]$ indicates flow de-collimation. 
In the following sections, we describe the effects of various physical parameters that
could affect the dynamics of the outflow.

\subsubsection{Radiation field and magnetic field strength} 
Here we discuss the interrelation between radiative forces and a variation of magnetic 
flux and their combined effect on outflow collimation.
We compare simulation runs M30a055b1, M30a055b3, and M30a055b5, all assuming the same stellar 
mass $30 \msun$, an inner jet launching radius of 1\,AU, and a density at the base of the flow of
$5.0\times10^{-14} {\rm g\,cm^{-3}}$, while the initial magnetic field strength parametrized 
by $\beta_0$ ranges from 5.1\,G to 11.5\,G.

The (initial) pure MHD runs exhibit characteristic differences. 
A lower $\beta_0$ provides faster jets which are more collimated, simply because the larger 
field strength provides larger Lorentz forces to collimate and also accelerate the flow more 
efficiently. 
A lower field strength implies a lower magneto-sonic wave speed, resulting in magneto-sonic 
surfaces located closer to the disk surface. 
Thus, we cannot simply compare the opening angles at the critical
surfaces as their positions vary in the pure MHD runs with different field
strengths.
Instead, we quantify the actual {\em change of opening angle} with and without considering 
radiative force $\Delta \zeta [\%]$ (i.e. in percentage), as profile along the field line.

\begin{figure}
\centering 
\includegraphics[width=1.0\columnwidth]{\figurepath/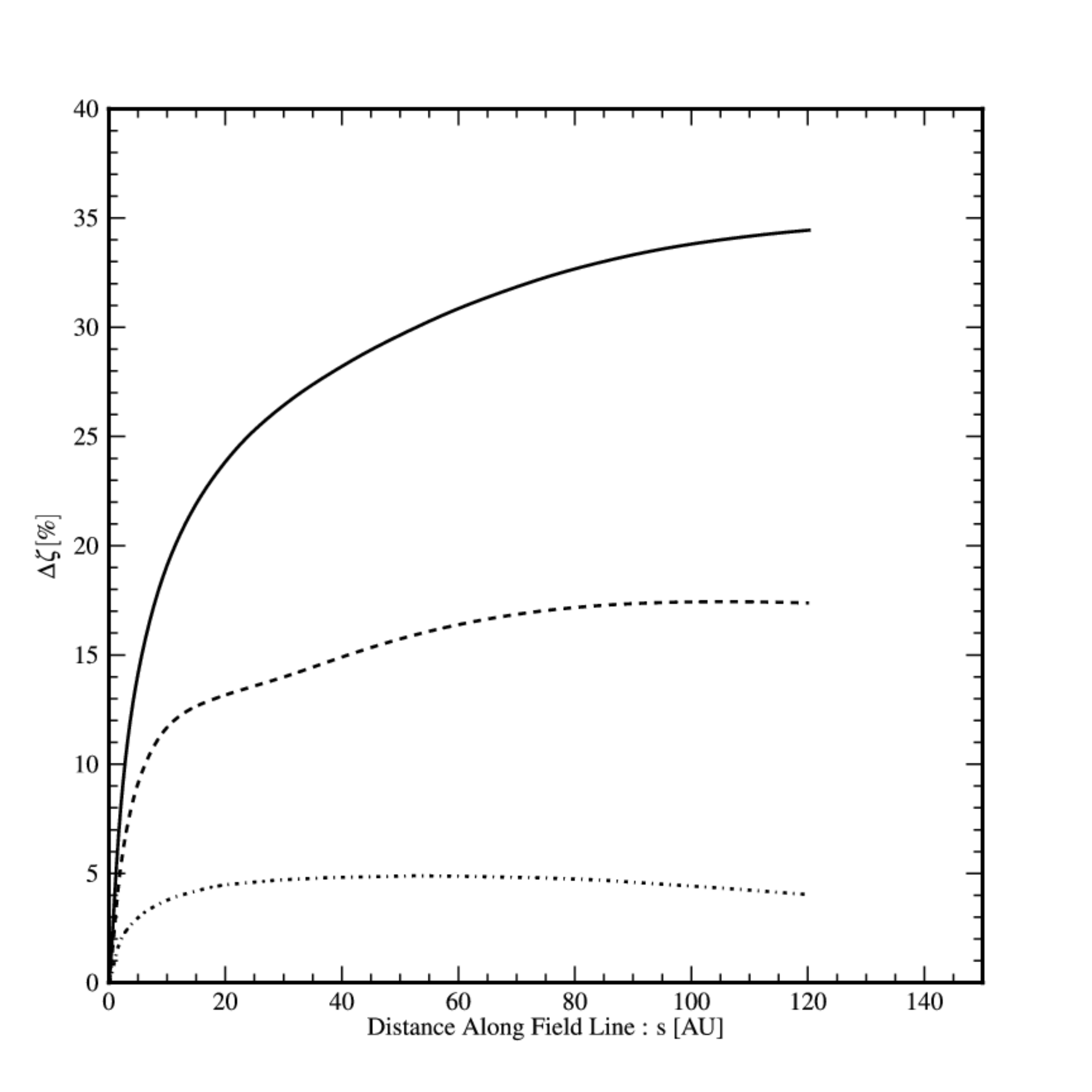}
\caption{Flow collimation along the field line rooted at r=5\, AU.
Shown is the profile of the percentage separation of field lines between the pure MHD flow and the radiative 
MHD flow [$\Delta \zeta [\%]$] as measure of flow collimation. 
The \textit{solid}, \textit{dashed} and \textit{dot dashed} lines are for three different 
plasma-$\beta_0 = 5.0, 3.0, 1.0$, respectively.}
\label{DiffBeta}

\end{figure}

This measure is shown in Fig.~\ref{DiffBeta} for a field line rooted at r = 5\,AU. 
We see that $\Delta \zeta [\%]$ is significantly positive, indicating that radiative 
forces do considerably de-collimate the flow. 
However, the effect of de-collimation is enhanced when the magnetic field strength is 
reduced (i.e. an increase of $\beta_0$). 
The asymptotic value of $\Delta \zeta [\%]$ along a field line changes from 35\% to 5\% 
when the initial magnetic field strength is increased by a factor of two.

The above analysis for different magnetic field strength suggests that in the strong field 
case ($\beta_0 = 1$) jet collimation is controlled by the magnetic forces alone. 
However, when we decrease the field strength (from 11.5\,G to 5.1\,G as normalized at 1\,AU)
the stellar radiation line-force begins to compete with the magnetic forces. 
The resulting jet has a much wider field structure and outflow opening angle as it is
de-collimated as compared to its pure MHD counterpart (see Fig.~\ref{evolvevz}). 
We conclude that radiative forces from the central star (for the chosen parameters)
will dominate the magnetic effects in controlling the flow collimation for field strengths
$\lesssim 5$\,G (at 1\,AU). In addition to the field strength the
  field profile is also important in dynamical evolution of the jet
  flow (see \S~\ref{ssec:jetregion}).
Observational estimates of the field strength in these close by regions of massive young stars 
will further help to narrow the results.
  
\begin{figure}
\centering
\includegraphics[width=1.0\columnwidth]{\figurepath/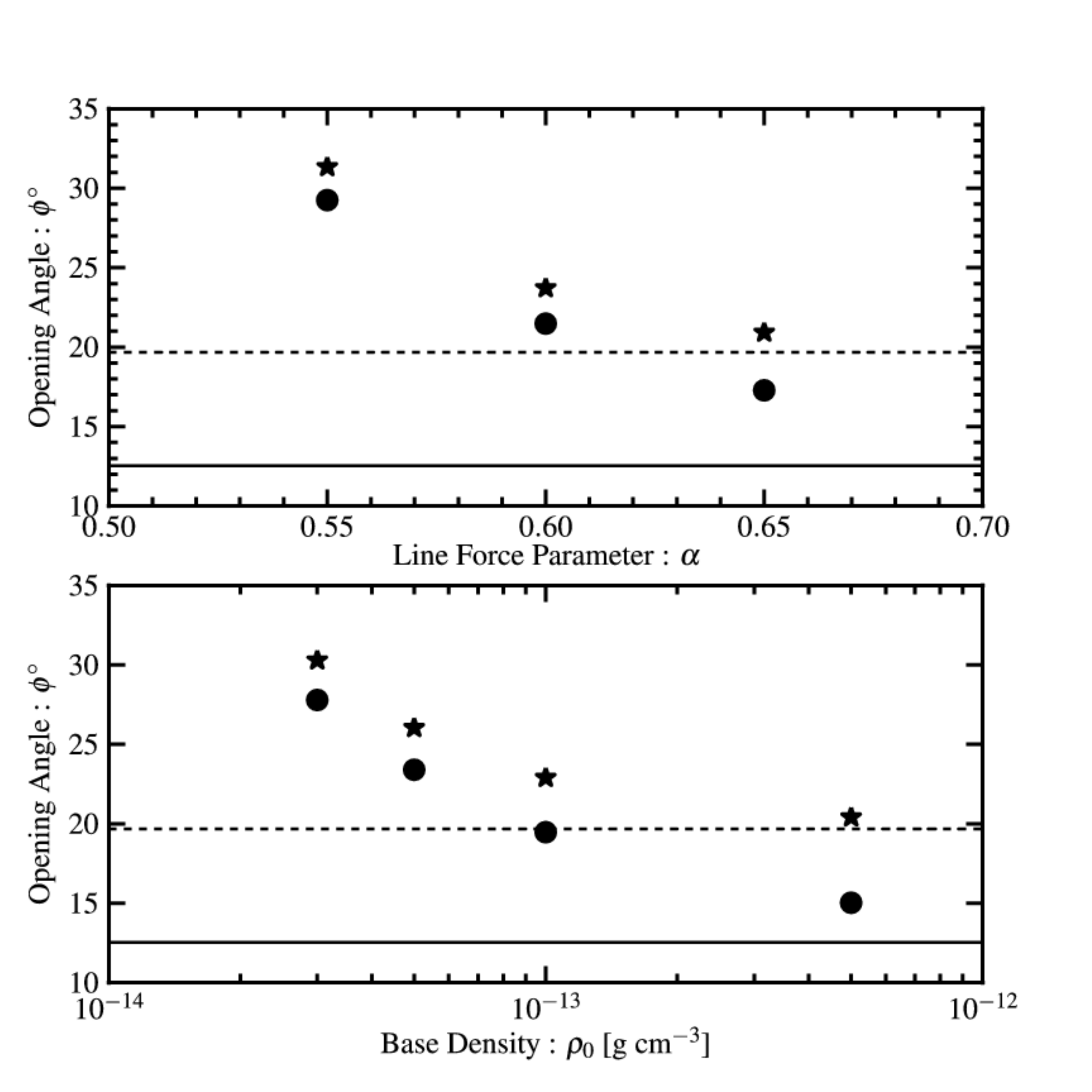}
\caption{Jet collimation, jet density and radiation force.
Shown is the opening angle (i.e. the field inclination) at the Alfv$\acute{e}$n point
(\textit{stars}) and the fast point (\textit{dots}) of the field line
rooted at r=5\, AU
for runs with different $\alpha$ (\textit{top panel}), and different jet base density 
$\rho_0$ (\textit{bottom panel}). The \textit{dotted} and \textit{solid} lines corresponds to 
the opening angle at the Alfv$\acute{e}$n point, and the fast point of the same field line 
after the pure MHD flow with an initial plasma-$\beta_0$ = 5.0.}
\label{DiffAlphaRho}

\end{figure}

\subsubsection{Impact of the line-force parameter $\alpha$}
The stellar line-force is a non-linear function 
of $\alpha$. 
\cite{Gayley:1995p1968} showed that a significant change in the mass flux can be obtained 
with very slight change in $\alpha$, indicating that this parameter is very sensitive for 
calculation of radiative line-forces. 

Simulation runs M60a055b5, M60a060b5, and M60a065b5 apply the same magnetic field strength
($\beta_0 = 5.0$), but differ in the line-force parameter from $\alpha = 0.55 - 0.65$.  
A first result is that the increase of $\alpha$ from 0.55 to 0.60 decreases the mass outflow rate
significantly by $\sim 16 \%$. We will discuss this interrelation in \S~\ref{ssec:injectmflux}.
However, also the outflow opening angles are affected as indicated by the top panel of 
Fig.~\ref{DiffAlphaRho}. 
The outflows with lower $\alpha$ become more de-collimated than the
others with higher $\alpha$.

The above trend indicates that the efficiency of the line driving
mechanism is increased with lower $\alpha$.
Physically a lower $\alpha$ implies a higher contribution of optically
thin lines in accelerating the flow. 
Thus, the dominance of less saturated (i.e. less
self-shadowed) lines in accelerating the flow results in an efficient line driving.

In summary, we find that the runs with lower $\alpha$ have higher mass flux and are 
less collimated. Our results from studying different $\alpha$ are
in qualitative agreement with the analytical results from \cite{Gayley:1995p1968}.

\subsubsection{Impact of the density $\rho_0$}
In this section, we describe how collimation and acceleration in the outflow are altered with 
the density at the base of the flow. 
As mentioned in \S~\ref{sec:rhobase}, the physical mass density applied to scale the numerical simulation 
is a free parameter in our setup, were we estimate its value from the observations using Eq.~\ref{valrhobase}.

Simulation run M30a055b5d1 has the lowest density among all four runs, 
$\rho_0=3\times10^{-14} {\rm g\,cm}^{-3}$  
For this run, the opening angles measured at critical MHD surfaces are larger compared 
to our reference run. 
Quantitatively, the opening angles at the Alfv$\acute{e}$n and the fast surface are 
$30^\circ$ and $28^\circ$, respectively. 
Also, the maximum percentage change in field line opening angle $\Delta\zeta[\%]$
increases from 35\% for the reference run to a staggering high value of 48\% for field line rooted at r = 5\, AU.
In case of a high jet base density $\rho_0$, the change in field line opening angle
$\Delta\zeta[\%]$ reduces to 4\%, 
with the opening angles at the Alfv$\acute{e}$n and fast surface being $20^\circ$ and $15^\circ$, respectively (Fig.~\ref{DiffAlphaRho}). 

These results correspond to an inverse correlation of the density at the base of 
the flow with the force multiplier
$M(\mathcal{T})$ (see Eq.~\ref{nondimforcemult}).
Higher densities result in an optically thick environment, thus increasing the 
the optical depth parameter $\mathcal{T}$, and reducing the force multiplier. 
Thus, the radiative {\em line-driving} force approaches the limit corresponding to the 
{\em continuum} radiation force. 
Essentially, the continuum force for typical massive young stars is weak compared to 
other dynamically important forces (for e.g. gravity).

In summary, we observe the outflow density as one of the leading 
parameters to govern the dynamics of the outflow by radiation forces. 
However, observationally this density is very difficult, even impossible to determine. 
We thus have to rely on certain assumptions, or may apply estimates of the outflow flux 
to calculate this density. 
Our parameter survey considers density values which are consistent with the observed 
mass fluxes. 
From our studies, we observe in general that for densities $\rho_0 < 10^{-13} {\rm g\,cm}^{-3}$ 
the line-driving force from the central star is very efficient in accelerating and 
de-collimating the outflow. 
However, a denser environment will dilute the influence of the line-driving mechanism. 
\subsubsection{Stellar mass evolution and outflow collimation}
\label{sssec:masssurvey}
\begin{figure}
\centering
\includegraphics[width=1.0\columnwidth]{\figurepath/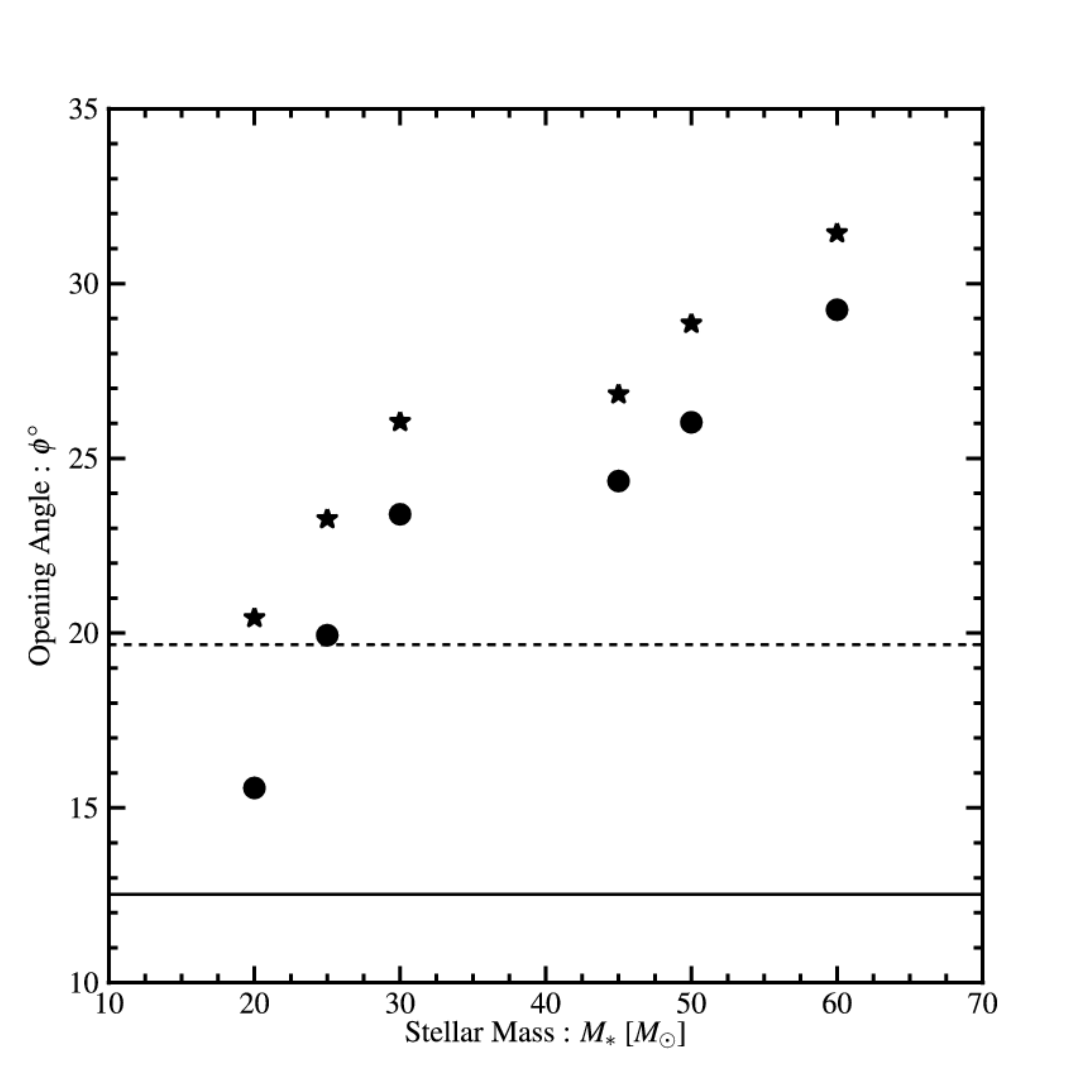}
\caption{Same as figure~\ref{DiffAlphaRho}, but for simulation runs for different stellar mass.}
\label{DiffMass}

\end{figure}

Motivated by the hypothesis of \cite{Beuther:2005p791} we have studied the outflow 
dynamics and collimation
for a sequence of different stellar masses (viz. from $20\,\msun$ to $60\,\msun$). 
The change in stellar mass implies a change in the central luminosity.
Fig.~\ref{DiffMass} shows the variation of the opening angle at the MHD critical points 
in simulation runs with central stellar mass. 
It can be seen that the opening angle varies from 
$20^\circ$ to $32^\circ$ for stellar masses from $20\,\msun$ to $60\,\msun$. 
The curve rises linearly up $30\,\msun$, but then the opening angle 
does not change considerably for increasing mass. 

For the physical parameters of the stellar evolution such as stellar luminosity and radius,
we have applied the {\em bloating star} model of \citet{Hosokawa:2009p4005} assuming an
accretion rate of $1.0\times 10^{-3}\msun {\rm yr}^{-1}$ (see fig.~\ref{hosogammae}). 
In this model for massive young stars, the stellar luminosity increases rapidly from $6\,\msun$ 
to $30\,\msun$. 
A star with high accretion rate undergoes swelling and is considered to bloat up to $100\,\rsun$. 
It is the entropy distribution in these stars which causes them to expand in size that much. 
For stars $< 10\,\msun$, a thin outer layer absorbs most of the entropy from its deep interiors, 
thereby increasing the entropy in this layers. 
The rapid increase of entropy in the outer layer of the star causes the star to increase in 
size \citep{Hosokawa:2009p4005}.
When the stellar mass reaches $\sim 10\,\msun$, the star then begins to shrink (Kelvin-Helmholtz
contraction) until its mass reaches $\sim 30\,\msun$. 
After that, the star approaches the main sequence and then follows the typical main sequence 
evolution. 
The stellar luminosity does not change considerably from 
$M_{*} = 30\,\msun$ to $40\,\msun$ - in fact there is a slight decrease. 
Contrary, for stars with mass $>40\,\msun$ the stellar luminosity increases with mass. 
This is reflected as a slight dip in the variation of $\Gamma_{\rm e}$ with mass between 
$30\,\msun$ and $45\,\msun$. 
Thus, the values of $\Gamma_{\rm e}$ derived for these two masses do not show considerable
difference (0.2381 and 0.2269 respectively). 
We conclude that the radiation force and, hence, also the dynamics of 
the outflow do not change to a great extent for these masses.

For simplicity we keep the inner jet launching radius and the density at the base of the 
flow fixed for all mass runs. The radiation force is not only altered by changes in $\Gamma_{\rm e}$,
but also due to the physical scaling that appears in the prescription of the force multiplier. 
In particular, this is the Keplerian velocity at the inner launching radius $l_0$, 
which is naturally different for different masses and scales with $\sqrt{M_{*}}$. 
One would then expect that the radiative force changes by a factor equivalent to $\sqrt{M_{*}}$. 
The effect of such a change with increasing central mass not only enhances the mass flux launched 
from the disk boundary but also imparts the resulting outflow a wider morphology (see Table~\ref{VaryB}).

In summary, with the above parameter study of different central stellar masses, we observe that 
as the luminosity (or mass) of the central star increases, the stellar radiation force becomes 
relatively stronger.
This leads to a higher degree of outflow de-collimation confirming the above mentioned
observational picture of outflow evolution in massive young stars.

\subsection{Radiation field and jet mass flux}
\label{ssec:injectmflux}
The simulations presented so far have been performed with "floating" boundary conditions
at the base of the flow (see \S~\ref{ssec:boundary}).  
While guaranteeing a causally correct boundary condition \textit{ab initio}, the
disadvantage is that the mass flux emerging from the underlying disk boundary cannot be
prescribed \textit{a priori}.
In fact, the mass flux is self-consistently calculated each time step by ensuring continuity 
of outgoing waves between the domain and the ghost cells. 
Thus, in our approach applied so far we fix the initial flow density and float the vertical outflow
velocity at the boundary. 

\begin{figure} 
   \centering
   \includegraphics[width=1.0\columnwidth]{\figurepath/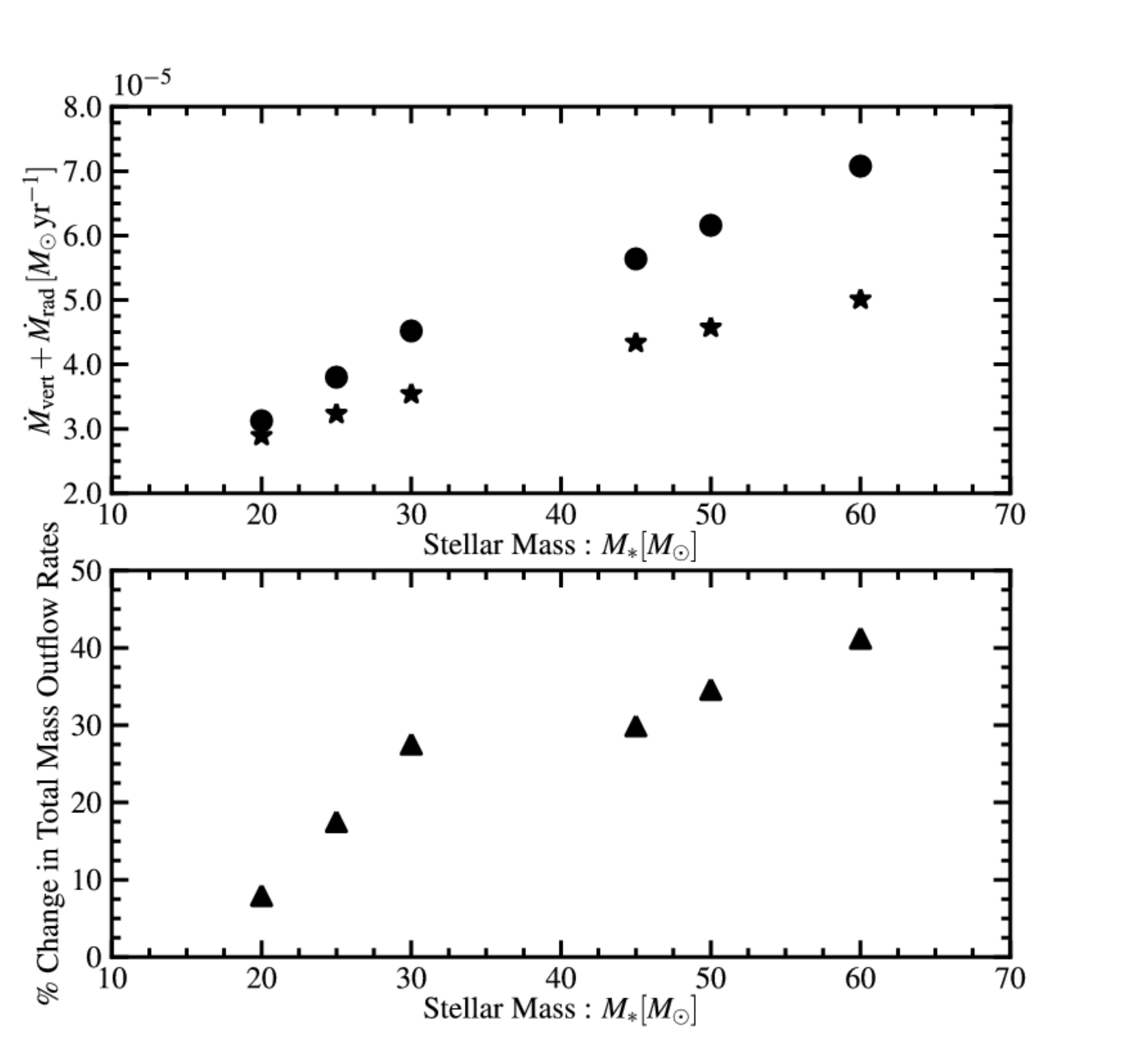}
\caption{Outflow mass flux for simulations applying a different stellar mass. 
   The \textit{top panel} shows the total mass outflow rates for each star in case of a pure MHD 
   flow (\textit{stars}), and for the same outflows including radiative forces (\textit{dots}). 
   The percentage difference between these mass fluxes is shown in the \textit{bottom panel}.}
   \label{mflux_flt}
\end{figure}

The total mass outflow rate ($\dot{M}_{\rm rad} + \dot{M}_{\rm vert}$) in physical units 
is shown in Fig.~\ref{mflux_flt} for runs with different stellar mass and applying
a floating boundary condition. 
The total mass flux has a $\sqrt{{\rm M}_{*}}$ dependence for
pure MHD runs on converting it to physical units.
While for a steady-state radiative MHD flow, the physical mass flux also 
depends on the Eddington parameter $\Gamma_{\rm e}$, which is related to the stellar 
luminosity. 

The percentage change between these mass outflow rates is shown in the bottom panel of
Fig.~\ref{mflux_flt}. 
Interestingly, the profile of this curves is similar to the variation of the opening 
angles (see Fig.~\ref{DiffMass}),
indicating that the increased mass flux could in fact play a role for de-collimation. 
Thus, the combination of both the "floating" boundary conditions
and the disturbance of the gas physics at the base of the flow by radiative source terms 
leads to the enhanced mass flux, which modifies the collimation of the flow.  

However, also the direct effect of radiative force on the flow can physically deflect 
the flow and eventually lead to de-collimation. 
Thus, the problem becomes quite complex as the direct influence of radiative force 
on the flow is mingled with its second order effects for e.g. increase in the mass flux. 
In order to single out the influence of direct de-collimation effects by radiative forces, 
we have performed simulations where we have fixed the outflow mass flux as a boundary condition.

\begin{figure} 
   \centering
   \includegraphics[width=1.0\columnwidth]{\figurepath/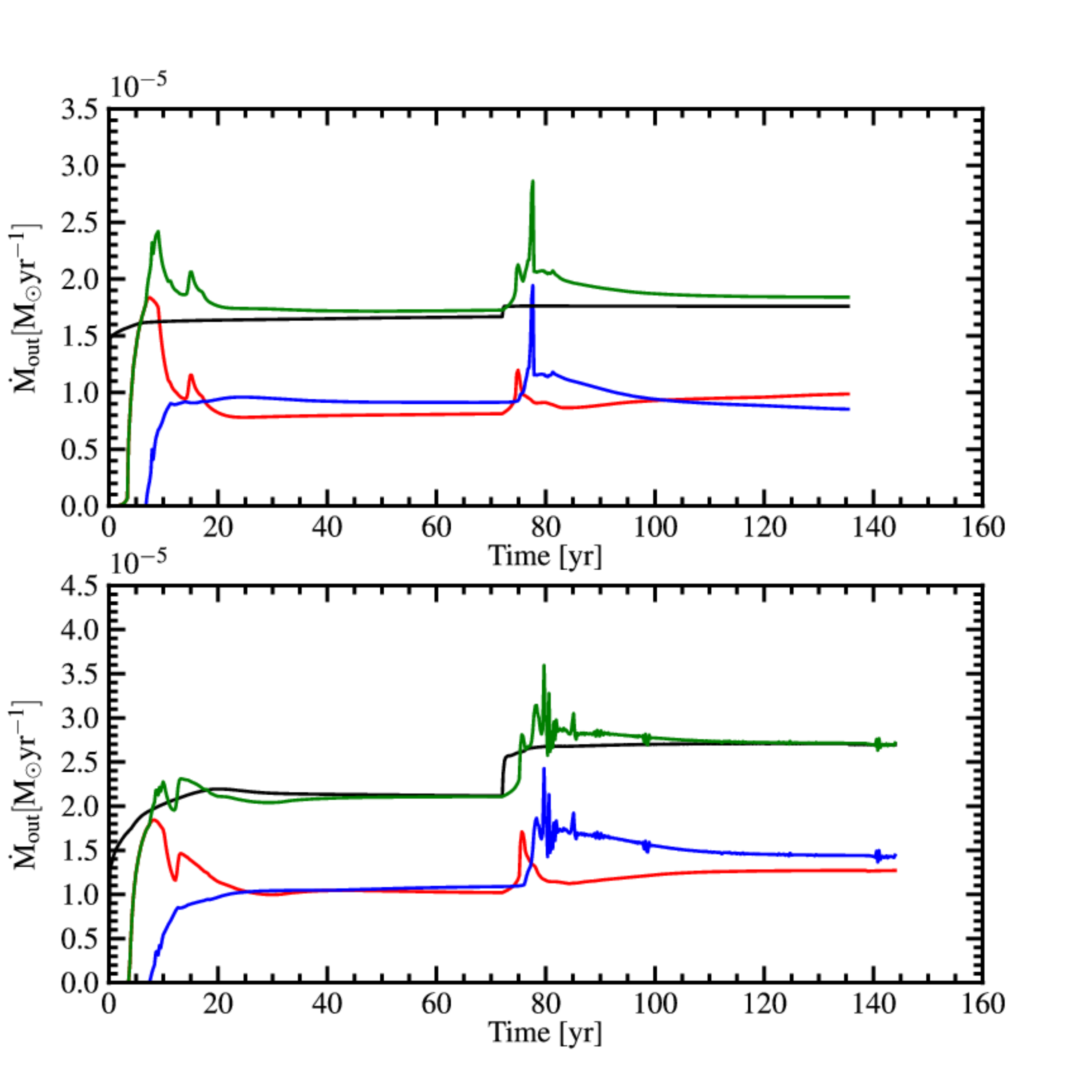}

   \caption{The variation of the outflow mass flux with time in simulation runs M30a055b3inj (\textit{top panel}) 
    and M30a055b5 (\textit{bottom panel}). 
    The \textit{red} and \textit{blue lines} show the variation of the vertical and radial mass fluxes, 
    respectively (see Eqs.~\ref{radmout},\ref{vertmout}). 
    The \textit{black line} corresponds to the mass flux injected from the underlying disk.
    The \textit{green line} is the total mass flux leaving the
    computational domain. Coinciding green and black lines proves
    conservation of the mass flux.}
   \label{mflux_inj}
\end{figure}

It is essential to inject the outflow with a velocity which is supersonic to begin with.
We fix the mass flux at the boundary by setting the inflow 
velocity to $v_{\rm z} = 0.24 \times v_{\rm Kep}$,
resulting in a flow with slow magneto-sonic Mach number close to unity and 
indicating a slightly supersonic flow. 
In Fig.~\ref{mflux_inj} we show the mass outflow flow rates derived along the top 
($z = z_{\rm m}$) and right ($r = r_{\rm m}$) boundaries (in red and blue, respectively). 
We have calculated the radial and vertical mass  outflow rates using following 
expressions,
\begin{equation}
\label{radmout}
\dot{M}_{\rm rad} = r_{\rm m} \int_{0}^{z_{\rm m}} \rho v_{r}  |_{r_{\rm m}} dz,
\end{equation}
\begin{equation}
\label{vertmout}
\dot{M}_{\rm vert} = \int_{0}^{r_{m}} r \rho v_{\rm z}  |_{z_{\rm m}} dr.
\end{equation}

In the simulations with floating boundary conditions, we observe that the mass flow rate 
reaches a steady-state, first derived self-consistently in the pure MHD flow, 
and then suddenly increased on adding the radiative source term.  
Essentially, this rise is not seen when we apply fixed-mass-flux boundary conditions. 
However, the fact that the curves of radial and vertical mass flux intersect 
after $\sim 60$ disk rotations after switching-on the radiation force, 
is a clear signature of de-collimation. 
Thus, the modification in terms of collimation due to the {\em direct} impact of
radiative forces on a flow injected with constant mass flux can be now understood 
with our simulation runs with {\em a priori} fixed-mass-flux boundary conditions.

We have thus performed three simulation runs with different stellar mass applying the 
above-mentioned mass-injection boundary condition (see Table~\ref{VaryB}). 
The most evident measure of de-collimation that can be observed is the 
ratio of vertical to radial mass flow rate. 
A higher ratio indicates a higher degree of jet collimation. 
This mass flux ratio for the chosen stellar masses of 25, 30, and 50\,$\msun$ 
changes from 1.12 after the steady-state pure MHD flow to 0.92, 0.85, and 0.80 
respectively, when radiative force is considered. 
Note that the radiative forces also accelerate the flow, increasing the maximum velocity
for M30a055b3inj in steady-state radiative MHD by a factor of 
$\sim 1.5$ compared to pure MHD flow.

We also find, that de-collimation is mainly due to the {\em direct} influence 
of the stellar line-driven force on the outflow,
as the amount of mass flux launched into the domain is fixed for these runs
(it is thus not a 2nd-order effect resulting from a higher mass flux due 
triggered by the radiation force).

In summary, we conclude here that the line-driving radiation force from the star 
plays a significant role in directly modifying the dynamics of previously launched MHD flow. 
This force not only aids in acceleration of the flow, but also pushes the outflow 
material away from the axis resulting in a substantially wider opening angle.  

\subsection{Acceleration by the disk radiation field}
 We have also carried out a number of simulations that involve only 
 radiation forces caused by a (hot) accretion disk.
In principle, in addition to the intrinsic accretion luminosity, the disk radiative 
 force can be enhanced by considering irradiation from the central star 
 (\citealt{Proga:1998p854}, \citealt{Drew:1998p876}). 
For the present model, however, we do not consider irradiation.

Here we present results of two of our simulations, denoted as Disk1 and Disk2 
(see Table~\ref{Diskandinj}). 
We have applied the parameter set as for the reference simulation M30a055b5,
however, two more parameters governing the dimensionless accretion rate that 
appears in the accretion luminosity (Eq.~\ref{nondimMacc}) have to be prescribed.
That is the ratio $\mu$ of disk to stellar luminosity, 
and the ratio $\Lambda$ of stellar radius to inner launching radius
(see Table~\ref{ParaSummary}). 
Naturally, a disk extending to radii closer to the center of gravity
would have much higher temperatures thus luminosity 
(see Vaidya et al. 2009 for an application to massive young stars) 
and the resulting radiation forces would be able to affect the outflow 
more.

For the number values applied for these additional parameters we again follow 
the {\em bloating star} model \citep{Hosokawa:2009p4005} assuming an accretion 
rate of $10^{-3}\msun {\rm yr}^{-1}$, and an inner jet launching radius
of $l_0 =0.1\,\rm{AU}$, thus, obtaining $\mu = 0.4644$ and $\Lambda = 0.4969$.
Note that the smaller inner jet launching radius also implies a inner disk radius
closer to the star, compared to our models discussed above.

For simulation run Disk2, we have increased the magnitude of the disk radiative 
force by decreasing the density at the inner jet launching radius by a factor 
of 10, thus assuming $\rho_0$ = $5\times10^{-15}{\rm g\,cm}^{-3}$.
As the radiation force is very sensitive to the density, 
the lower density at the jet base does increase the radiation force from the disk
by a factor of three. 
Note that, physically, this increase of the disk radiation force could mimic
the effects of stellar irradiation.

We quantify the change in collimation degree in runs Disk1 and Disk2 again with the parameter
$\Delta\zeta[\%]$ (see Eq.~\ref{delzeta}). 
Simulation run Disk1 has a maximum of $\Delta\zeta[\%] = -1.8$, while run Disk2 has 
$\Delta\zeta[\%]= 3.1\%$. 
These number values are rather low compared to our simulations with a stellar radiation
force indicating that the
{\em{disk radiation force alone is not strong enough to de-collimate the flow}}. 

Interestingly, simulation run Disk2 shows some unsteady behavior close to the axis - a feature
which is absent when only stellar radiation forces were considered.
This is consistent with the findings of \cite{Proga:1999p872}, who detect a similarly unsteady 
behavior in their simulations for the case when the radiative force is dominated by the 
underlying disk. 
The fact that we do not see this feature in simulation Disk1 suggests that the radiative 
force in run Disk1 is not comparable to the other forces that control the flow dynamics,
and that this outflow does more correspond to a pure MHD flow. 
The unsteady flow behavior in run Disk2 is also reflected in the poloidal velocity 
evolution as steadily propagating 'wiggles' in the velocity profile along the field line
(or streamline) $v_{\rm p} (s)$ (Figure \ref{ldinstablefig}). 

As maximum outflow velocity for the simulations Disk1 and Disk2 we obtain
relatively high velocities of $\sim$ 350 kms$^{-1}$.
This is mainly due to the above mentioned fact that the outflow is launched 
deeper in the potential well and as close as $l_0$ = 0.1\,AU from the central 
star.
Compared to the pure MHD run, we see, however, that the jet which is affected by
disk radiation forces (only) achieves slightly higher asymptotic velocities, as 
the MHD reaches only $\sim$ 300 km s$^{-1}$.
Thus, the disk radiation force affects the outflow primarily by slightly 
accelerating it. 
We do not see indication that the outflow collimation is affected,
which is understandable since the disk radiation force acts mainly
in vertical direction.

In summary, jet acceleration and collimation is rather weakly affected 
by the disk radiative forces as their magnitude is orders of magnitude 
smaller compared to radiation forces from the central star. 
However, for other astrophysical jet sources such as CVs and AGNs, the radiative 
force from the disk could play a significant role. Applying the scaling relations and comparing the amount
of energy radiated per unit area from the standard thin disk from a
typical young massive star (MS) and a standard white dwarf (WD) we
obtain from equation~\ref{disk_energy}, 
\begin{equation}
\frac{D^{MS}}{D^{WD}} = 0.06
\left(\frac{10^{-3}\,\msun\,\rm{yr}^{-1}}{10^{16}\rm
  g\,s^{-1}}\right)\left(\frac{30\,\msun}{1\,\msun}\right)\left(\frac{0.1\,\rm
    AU}{10^{9}\,\rm cm}\right)^{-3}
\end{equation}. 
This indicates that the efficiency of line driving due to disk force
alone for such massive proto-stellar disks is less than the compact
object accretion disks like CVs which have been the focus of previous
studies by \cite{Pereyra:2003p1144}. This ratio is even smaller if we
consider disks around AGNs as considered by \cite{Proga:2004p838}.

\subsection{Limitations of our model approach}
\label{ssec:limitations}
The main goal of our study was to disentangle the effects of radiative forces
from the young star-disk system on a pure MHD outflow launched within the
standard picture of magneto-centrifugal, magnetohydrodynamic jet formation.
In this section we discuss the limitations of our model approach to the subject
of jet formation from massive young stars.

\subsubsection{Prescription of radiation force}
The prescription of radiation force used for the present model does not 
explicitly consider the radiation transfer.
Instead it implements the radiative force due to lines as a source term in  
the momentum and the energy equations (see \S~\ref{sec:eqns}). 
This greatly simplifies the numerical task and modeling becomes computationally 
inexpensive.

\subsubsection{Possible existence of a stellar wind}
Our model does not consider a stellar wind from the massive young star.
Observed outflows around young massive stars are usually thought to have velocities 
of about 200 - 500 km\,s$^{-1}$, a value much smaller than the speed of stellar winds 
in the evolved stages of OB stars (of about $\sim$ 1000 km\,s$^{-1}$). 
This  difference in velocity scale could give a clue to difference in  
environment around the star. 
During the formation stage, the star is surrounded by  dense gas and 
dust as compared to more evolved stage where it has cleared all 
surrounding matter. 
A rarer environment in a more evolved stage could lead to an efficient 
acceleration of winds via line driving to  1000 km\,s$^{-1}$. 
Hence, for more massive and hotter young stars, impact due stellar winds 
could play a vital role on the dynamical evolution  of disk winds. 

\subsubsection{Lack of knowledge of system parameters}
The inner regions around massive young stars are not accesible with present 
day telescopes. 
Thus, many physical parameters for these inner regions are not observationally 
constraint (for e.g\, the inner disk radius, mass density, magnetic
field strength) 
In the present model, we follow  the notion that high-mass stars form as a 
scaled-up low-mass stars.  
Thus most of the parameters used for the present modeling are derived from 
estimates for low-mass stars. 

\subsubsection{Ideal MHD and line force parameters}
Simulations presented here are done using ideal MHD approximation.
This is in principle fine as ionizations fractions are usually high
enough to provide a good coupling between matter and field.
However, for simplicity we have assumed a constant ionization fraction 
throughout the wind. 
The radiative force parameters do have an explicit dependence on
the ionization fraction. The varying temperature across the jet implies to
 varying degree of ionization which would modify the line force
  parameters and thus the value of the force. In this model, we use an idealized assumption of a constant ionization 
 fraction throughout and fix the radiative force parameters for all
 simulation runs. While in case of non-magnetic CVs,
 \cite{Pereyra:2003p1144} has shown that the qualitative effect
 of the line force is not altered by incorporating local ionization
 effects for the line force parameters. However, the importance of
 such local ionization effects for the case of young massive stars is
 not known in the literature.

\subsubsection{Prescription of the disk dynamics}
A self-consistent modeling would need to incorporate the time-evolution
of the disk structure in the simulation, and to treat accretion and ejection
processes simultaneously. 
Such simulations are currently performed in general applications of jet launching
(e.g\, \citealt{Casse:2002p1142}, \citealt{Zanni:2007}, \citealt{Murphy:2010p6874}), however, it is too early to be applicable to massive
young stars - one reason being the lack of knowledge of the accretion disk 
parameters (e.g. the question whether there is a thin or thick disk), another
one that also radiative effects would have to be implemented in these simulations.

%
%
\section{Conclusions}
\label{conclusions}
We have studied the impact of a line-driven radiation forces on the acceleration and collimation
of a MHD jets,around young massive 
stellar object.
Our main motivation was to give a solid theoretical understanding for the outflow evolution hypothesis
presented by \cite{Beuther:2005p791}.
For the radiation forces we have considered stellar and disk luminosity.
Our basic approach was to initially launch a MHD jet from the underlying disk surface, and then, 
after the pure MHD outflow has achieved a steady state, to switch on the radiative forces and study 
their effect on the existing MHD outflow.

We performed a number of simulations with the line driving force exerted by 
stellar radiation. 
These simulations were performed for a wide range of physical parameters, as 
 i) the central stellar mass, 
ii) the magnetic flux, and 
iii) the line-force parameter $\alpha$. 
In order to apply our method of calculating the line driving force (CAK approach), 
we have modified the numerical code PLUTO to incorporate appropriate
projections of gradients of the 2\,D velocity field along the light path.
Additionally, we have consistently implemented these projections for different
radiation sources properly considering the geometry of the radiation
field. All these simulations have 'floating' and casually
  consistent inflow boundary in
which the mass flux is not fixed and is derived by the
numerical solution.
  
Our main conclusions from this analysis are as follows.

The line driven force from the central star for the parameters considered 
does play a significant role in modifying the dynamics in terms of collimation and acceleration of the outflow. 
We find that the outflow velocity is increased by a factor of 1.5 - 2 by radiation forces
as compared to the pure MHD flow. 
Also, the degree of collimation is lowered, visible e.g. in a 30\%-change in the magnetic flux profile, or
the wider opening angle of the magnetic field lines.

Investigating different stellar masses we determine the amount of de-collimation by measuring 
the opening angles of a typical field lines (that with the highest mass flux) at the 
Aflv$\acute{e}$n and fast magneto-sonic point. 
We find that for a stellar mass of $20 \msun$, the opening angles are $20^\circ$ and $16^\circ$,
respectively.
For a $60 \msun$ star these values increase to $32^\circ$ and $29^\circ$, indicating substantial amount
of de-collimation due to the increased stellar luminosity.
This findings confirm the observed evolutionary picture for massive outflows in which more massive stars tend to
have outflows of lower collimation degree.
Note that for massive young stars, our results do not only constitute a relation of different stellar masses,
but correspond also to a time-evolution of outflow systems, as the central mass can be substantially increased 
during massive star formation.

We have also performed simulations with injection of fixed mass flux from the disk
boundary for various stellar masses. We find that the ratio of vertical to radial mass flux in
these runs decreases from 0.92 to 0.80 with increase in stellar mass
from $25\,\msun$ to $50\,\msun$. This clearly indicates the fact that the line driving
force from central star plays an influential role in the physical
understanding of observed evolutionary picture pertaining to outflows
from young massive stars.

So far, the magnetic field strength in the jet formation region close-by a massive young star forming is an 
unknown quantity. 
We therefore have carried out simulations with different field strength (plasma-$\beta_0 \sim 1.0, 3.0, 5.0$).
We find that for the flow with high magnetic flux the radiative forces do not really affect the
collimation degree. 
However, for the flow with low magnetic flux, the dynamical effect of radiative forces is greatly 
increased. 

We further find that the line force parameter $\alpha$ is very critical in determining the magnitude of the line-driven 
forces. 
Lower values of $\alpha$ leads to an efficient radiative force from the central star and thus decollimate the flow 
to a larger extent as compared to higher $\alpha$ values. 
Since the radiation forces also affect the mass outflow rates for simulations, even small change in $\alpha$ may lead
to significant changes in mass flux up to $\sim 28 \%$.

Line forces due to the hot accretion disk do not play a significant role in controlling the 
dynamics of the MHD outflow, simply because they are orders of magnitude smaller then all 
other forces that affect the flow dynamically.
Implementing high disk radiation forces has (by taking into account the inner hotter
part of the disk), however, shown us that the disk radiation forces will affect
primarily the flow acceleration, and not the flow collimation.

The source terms for the line-driven forces from the star and the disk implemented in our simulations depends on certain 
scaling parameters.
We find that the physical scaling of the jet density is a leading parameter that affects the flow dynamics. 
Large densities make it difficult for the line-driving to act efficiently, resulting in a flow 
which is mostly dominated by MHD forces. 
However, less dense inner regions around massive young stars would allow efficient radiative line-driving, 
and thus accelerate and de-collimate the flow with great effect.

This paper has addressed a complex problem of jet
  launching from young massive stars. In doing so, we have applied a
  simplified prescription of the radiative force rather than a complete
  radiative transfer calculation. One limitation of our model is the lack of observational knowledge
  about various parameters particularly very close to the central star. In addition, we
  have not included effects from stellar winds which might exist
  during the {\em{later stages}} of young massive star
  evolution. In spite of these limitations, we find clear evidence of
  acceleration and de-collimation of jets launched from massive
  young stars.

In summary, among all the radiative sources considered to study the dynamics of outflow launched from the young massive
star, we see that the line force from the central luminous star is very efficient in de-collimating and accelerating the 
flow. 
The line force from the underlying disk is not as significant as compared to the stellar force. 
Also, dynamical effects on the outflow due to the optically thin
electron scattering continuum force from the central
star and the disk is not significant. 
Furthermore, we confirm the observed trend of de-collimation seen in outflows from massive stars at different
evolutionary stages.

\begin{acknowledgements}
We acknowledge the Klaus Tschira Stiftung for funding this work
carried out at Max Planck Institute of Astronomy, Heidelberg and
also convey our thanks to the Heidelberg Graduate school of
Fundamental Physics (HGSFP). We appreciate insightful comments 
from Daniel Proga on this work and we thank him for his constructive suggestions. We also like to thank A.
Feldmeier for his helpful comments for
this work. We also forward our thanks to T. Hosokawa for providing his
data for the stellar evolution model considered here.
\end{acknowledgements}

%
%
\appendix
\section{Basics of Castor, Abbott and Klein Theory}
\label{sec:basicsCAK}
According to the Castor, Abbott and Klein theory (\citealt{Castor:1975p898}), the force multiplier can be expressed as,
\begin{equation}
 M(\mathcal{T}) = \sum_{\rm lines} \left(\frac{\Delta \nu_{\rm D} F_{\nu}}{F} \frac{1 - e^{-\eta \mathcal{T}}}{\mathcal{T}}\right),
\end{equation}
where
$\Delta \nu_{\rm D}$ the Doppler shift,
$F_{\nu}$ the radiation flux at frequency $\nu$, and 
$F$ the total integrated flux. 
The optical depth parameter $\mathcal{T}$ is related to the gradient of velocity, the density in the wind
and the ion thermal velocity $v_{\rm th}$,
\begin{equation}
\label{eq:tform}
 \mathcal{T} = \frac{\rho \sigma_{e} v_{\rm{th}}}{|\hat{n}\cdot\nabla (\hat{n}\cdot\vec{v})|}.
\end{equation}
The optical depth parameter $\mathcal{T}$ can be related to the optical depth of a particular 
line $\tau_{\rm L}(\hat{n}) = \eta \mathcal{T}$.
The \textit{line strength} $\eta$ is the ratio of the line opacity $\kappa_{\rm L}$ to 
the electron scattering opacity $\sigma_{\rm e}$, while $\hat{n}$ is the unit vector along 
the line of sight (l.o.s.). 
The classification of lines in optically thin and thick lines is done
on the basis of interaction probabilities. 
Optically thick lines are those which have interaction probability of unity.
Lines with optical depths $\tau_{\rm L} < 1$ are optically thin, and their probability of 
interaction is $\tau_{\rm L}$.  
Based on this approximation, the force multiplier is separable and can be written differently
for optical depths very high or very low. 
When the gas is optically thick, $\tau_{\rm L} > 1$, the force multiplier depends only on 
the local dynamical quantities of the flow,
\begin{equation}
 M_{\rm thick}(\mathcal{T}) = \sum_{\rm thick lines} \left(\frac{\Delta \nu_{\rm D} F_{\nu}}{F} \frac{1}{\mathcal{T}}\right),
\end{equation}
while for the optically thin case $\tau_{L} < 1$ the force multiplier is independent 
of the local dynamics, but depends on the line strength of individual thin lines,
\begin{equation}
 M_{\rm thin}(\mathcal{T}) = \sum_{\rm thin lines} \left(\frac{\Delta \nu_{\rm D} F_{\nu}}{F} \eta \right).
\end{equation}

In general, for a gas distribution with a mixture of optically thick and thin lines, the 
empirical form of the total force multiplier integrated over all lines can be expressed 
as a power law, $M(\mathcal{T}) \sim k\mathcal{T}^{-\alpha}$, where $k$ and $\alpha$ are line force parameters.

\section{Line Driving force due to disk alone}
\label{sec:diskforce}
In addition to the line radiative forces from the central star, we also take into account
line forces from the underlying disk. 
The underlying disk cannot be considered as a point source 
but rather is an extended cylindrical source of radiation. 
Further, the disk luminosity varies with the radial distance from the central star. 
In the present simulations, we consider that the underlying disk as a steady-state standard 
thin disk with a temperature profile given by \cite{Shakura:1973p1000}.  
Thus, the energy radiated per unit area $D(r)$ at cylindrical radius $r$ is
\begin{equation}
\label{disk_energy}
D(r) = \frac{3 \dot{M}_{\rm{acc}} 
       G M_{*}}{8\pi r^{3}}\left[1 - \left(\frac{l_0}{r}\right)^{1/2}\right],
\end{equation}
where $\dot{M}_{\rm{acc}}$ is the steady-state accretion rate onto a central star of mass $M_{*}$. 
The inner launching radius is $l_0$. 
In case of line forces from the accretion disk, we calculate the radial and vertical 
radiation flux from the standard disk, $S_{\rm r}$ and $S_{\rm z}$, 
similar to \cite{Pereyra:2000p836}. 
Further, the l.o.s. velocity gradient in the force multiplier applied in case of disks is 
for simplicity reduced to $\partial v_{\rm z} / \partial {z}$, 
since the bulk of the radiation flux from the disk is in vertical direction,
\begin{equation}\label{flinedisk}
\vec{f}_{\rm line,disk} = \frac{\sigma_{\rm e}}{c}\left[S_{\rm r}
\vec{r} + S_{\rm z} \vec{z}\right] M(\mathcal{T}).
\end{equation}
Here, $S_{\rm r}$ and $S_{\rm z}$ are the radial and vertical flux components.
Both depend on the disk luminosity and are implemented in the code in their dimensionless 
form. 

In order to reduce to a dimensionless form, all the physical quantities
should be written as a product of its value in code units with
appropriate scale factor (see Eq.~\ref{scaledef}).  $S_{\rm r}$
and $S_{\rm z}$ are radial and vertical components of the flux emitted
from the disk surface.(\citealt{Hachiya:1998p1763},
\citealt{Pereyra:2000p836}) -
\begin{equation}
\label{Sr}
S_{\rm r} = \int^{2\pi}_{0}\int^{end}_{l_{o}}\frac{D(r')}{\pi}\frac{z^{\alpha + 1}(r - r\rm{cos}(\phi))}{\left[\left(r^2 + r'^2 + z^2 - 2rr'\rm{cos}(\phi)\right)^{1/2}\right]^{4+\alpha}}r'dr'd\phi,
\end{equation}

\begin{equation}
\label{Sz}
S_{\rm z} = \int^{2\pi}_{0}\int^{end}_{l_{o}}\frac{D(r')}{\pi}\frac{z^{\alpha + 2}}{\left[\left(r^2 + r'^2 + z^2 - 2rr'\rm{cos}(\phi)\right)^{1/2}\right]^{4+\alpha}}r'dr'd\phi,
\end{equation}
where D($\rm{r}^\prime$) is the amount of energy radiated per unit
area from the standard thin disk (Eq~\ref{disk_energy}). In the
present simulations, the accretion disk is treated as a boundary and
the accretion of matter in the disk is not considered. 
The mass accretion rate that appears in the above flux radiative components of the disk has to be prescribed on basis of dimensionless parameters.
\begin{equation}
\label{nondimMacc}
\dot{M}_{\rm acc} =\frac{4\pi c l_0}{\sigma_{\rm e}} \Gamma_{\rm e} \Lambda \mu. 
\end{equation}

The radiation flux given above has to be written in dimensionless form
to be incorporated in the simulations. These flux components in the
code units are given as follows -
\begin{eqnarray}
\label{Srdimless}
S_{\rm r,code} = \frac{3 c G M_{*}}{2\pi \sigma_{\rm e} l_0^2}\Gamma_{\rm e}\Lambda\mu\
     \int^{2\pi}_{0}\!\!\!\int^{end}_{l_0}\frac{1}{(r'_{\rm
    c})^{3}}\left(1 - \sqrt{\frac{1}{r'_{\rm c}} }\right)\nonumber \\
    \frac{z_{\rm c}^{\alpha + 1} (r_{\rm c} - r'_{\rm c}\rm{cos}(\phi))}{\left[\left((r_{\rm c})^2 + (r'_{\rm c})^2 +
    (z_{\rm c})^2 - 2r_{\rm c}r'_{\rm
      c}\rm{cos}(\phi)\right)^{1/2}\right]^{4+\alpha}} r'_{\rm c}dr'_{\rm c}d\phi,
\end{eqnarray}

\begin{eqnarray}
\label{Szdimless}
S_{\rm z,code} = \frac{3 c G M_{*} }{2\pi \sigma_{\rm e}
  l_0^2}\Gamma_{\rm
  e}\Lambda\mu\int^{2\pi}_{0}\!\!\!\int^{end}_{l_0}\frac{1}{(r'_{\rm
    c})^{3}}\left(1 - \sqrt{\frac{1}{r'_{\rm c}}}\right)\nonumber \\
\frac{z_{\rm c}^{\alpha + 2}}{\left[\left((r_{\rm c})^2 + (r'_{\rm c})^2 + (z_{\rm c})^2 - 2r_{\rm c}r'_{\rm c}\rm{cos}(\phi)\right)^{1/2}\right]^{4+\alpha}}
 r'_{\rm c}dr'_{\rm c}d\phi.
\end{eqnarray}

\begin{figure}
   \centering
    \label{disklf}
    \includegraphics[width=10cm]{\figurepath/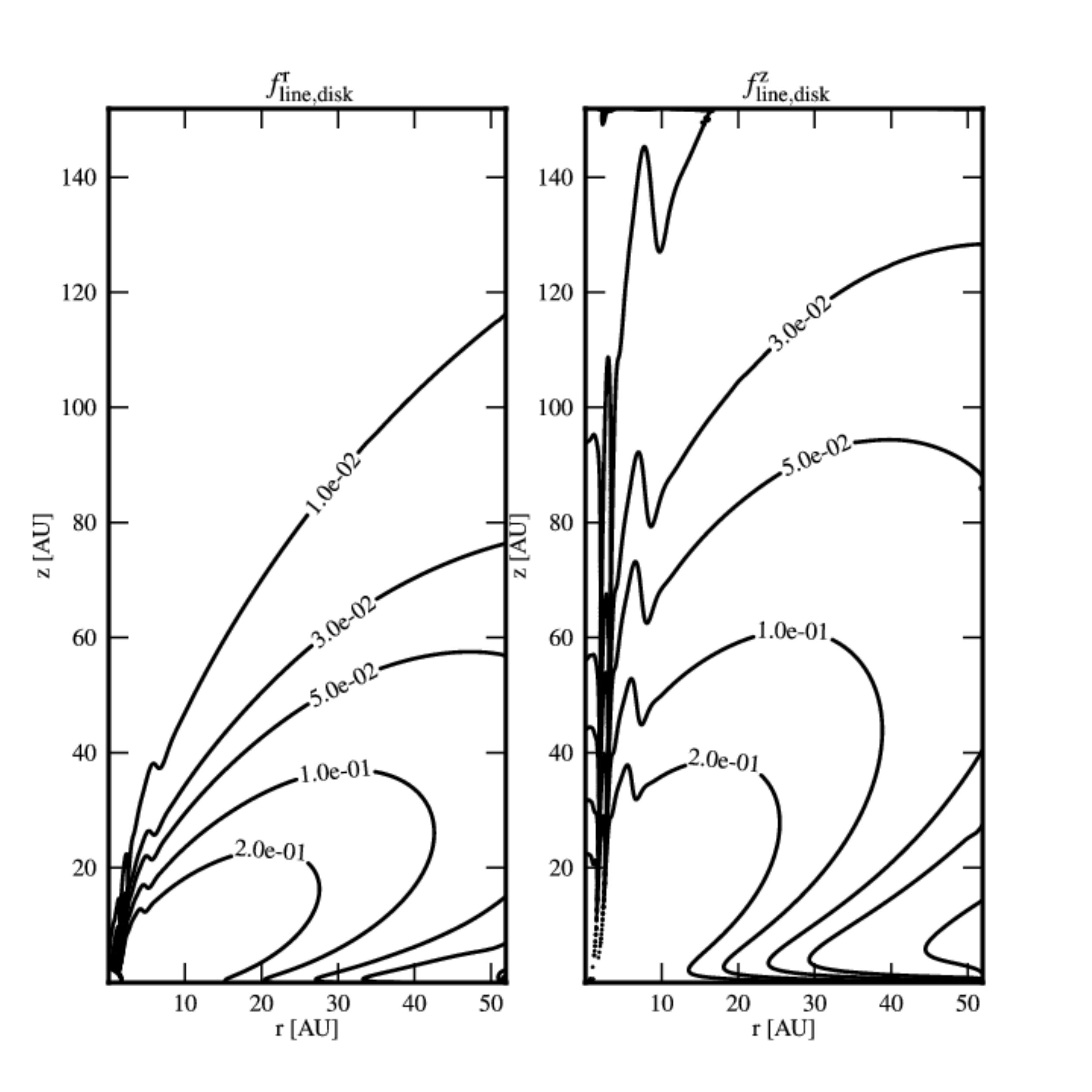}
   \caption{Contours of the line force from disk radiation in the radial (\textit{left panel}) and vertical
   (\textit{right panel}) direction. 
   The contour levels are given in physical units.
   The parameters used are : $Q_0 = 1400.0$, $\alpha = 0.55$, $M_{*} = 30 \msun$, $l_0= 0.1 AU$, 
   $\rho_0 = 5.0\times10^{-14}$ g\,cm$^{-3}$, $\Gamma_{\rm e} = 0.2369$, $\Lambda=0.4969$,
   $\mu = 0.4644$,$\beta_0 =5.0$ }
\end{figure}

Using the above formulations, the dimensionless radial and vertical  component of the line force from the disk 
can then be obtained. 
Their respective contours are shown in Fig.~\ref{disklf}.

\begin{equation}
f_{\rm{line,disk}}^{\rm{r,code}} (r_{\rm c},z_{\rm c}) = {f_{\rm{line,disk}}^{\rm{r}}}/(GM_{*}/l_0^{2}) = M_{\rm c}(\mathcal{T}) S_{\rm r,code},
\end{equation}

\begin{equation}
f_{\rm{line,disk}}^{\rm{z,code}} (r_{\rm c}, z_{\rm c}) =
{f_{\rm{line,disk}}^{\rm{z}}}/(GM_{*}/l_0^{2})= M_{\rm
  c}(\mathcal{T}) S_{\rm z,code}.
\end{equation}

The force multiplier for the case of disk force in code units is
similar to that used for stellar line force
(Eq~\ref{nondimforcemult}). However, since we assume that bulk of the photons
from the disk move along the vertical z axis, the line of sight
velocity gradient is approximated to be just due to vertical
velocity, 
\begin{equation}
\left|\frac{dv_{l}}{dl}\right| \sim
\left|\frac{dv_{z}}{dz}\right|.\nonumber
\end{equation}
The contours of the line force due to the underlying disk in the radial and the vertical
direction are shown in figure below.

\clearpage

\bibliographystyle{apj}


\end{document}